\documentclass[preprint2]{aastex62}

\def\kms{km s$^{-1}$}

\newcommand{\sbr}{mag arcsec$^{-2}$}
\newcommand{\solmass}{$M_\odot$}
\received{XXX}
\revised{YYY}
\accepted{January 2020}
\submitjournal{ApJ}

\shorttitle{KMTNet Nearby Galaxy Survey}
\shortauthors{BYUN et al.}

\begin{document}

\title{KMTNet Nearby Galaxy Survey II. Searching for Dwarf Galaxies in 
Deep and Wide-field Images of the NGC 1291 system}

\correspondingauthor{Minjin Kim}
\email{mkim@knu.ac.kr}

\author[0000-0002-7762-7712]{Woowon Byun}
\affiliation{Korea Astronomy and Space Science Institute, Daejeon 34055, Korea}
\affiliation{University of Science and Technology, Korea, Daejeon 34113, Korea}

\author[0000-0002-3211-9431]{Yun-Kyeong Sheen}
\affiliation{Korea Astronomy and Space Science Institute, Daejeon 34055, Korea}

\author[0000-0002-3505-3036]{Hong Soo Park}
\affiliation{Korea Astronomy and Space Science Institute, Daejeon 34055, Korea}

\author[0000-0001-6947-5846]{Luis C. Ho}
\affiliation{Kavli Institute for Astronomy and Astrophysics, Peking University, Beijing 100871, China}
\affiliation{Department of Astronomy, School of Physics, Peking University, Beijing 100871, China}

\author[0000-0003-3451-0925]{Joon Hyeop Lee}
\affiliation{Korea Astronomy and Space Science Institute, Daejeon 34055, Korea}
\affiliation{University of Science and Technology, Korea, Daejeon 34113, Korea}

\author[0000-0001-9670-1546]{Sang Chul Kim}
\affiliation{Korea Astronomy and Space Science Institute, Daejeon 34055, Korea}
\affiliation{University of Science and Technology, Korea, Daejeon 34113, Korea}

\author{Hyunjin Jeong}
\affiliation{Korea Astronomy and Space Science Institute, Daejeon 34055, Korea}

\author[0000-0002-6982-7722]{Byeong-Gon Park}
\affiliation{Korea Astronomy and Space Science Institute, Daejeon 34055, Korea}
\affiliation{University of Science and Technology, Korea, Daejeon 34113, Korea}

\author[0000-0001-9561-8134]{Kwang-Il Seon}
\affiliation{Korea Astronomy and Space Science Institute, Daejeon 34055, Korea}
\affiliation{University of Science and Technology, Korea, Daejeon 34113, Korea}

\author[0000-0002-6261-1531]{Youngdae Lee}
\affiliation{Department of Astronomy and Space Science, Chungnam National University, Daejeon 34134, Korea}

\author{Yongseok Lee}
\affiliation{Korea Astronomy and Space Science Institute, Daejeon 34055, Korea}
\affiliation{School of Space Research, Kyung Hee University, Yongin, Kyeonggi 17104, Korea}

\author{Sang-Mok Cha}
\affiliation{Korea Astronomy and Space Science Institute, Daejeon 34055, Korea}
\affiliation{School of Space Research, Kyung Hee University, Yongin, Kyeonggi 17104, Korea}

\author[0000-0002-9434-5936]{Jongwan Ko}
\affiliation{Korea Astronomy and Space Science Institute, Daejeon 34055, Korea}
\affiliation{University of Science and Technology, Korea, Daejeon 34113, Korea}

\author[0000-0002-3560-0781]{Minjin Kim}
\affiliation{Department of Astronomy and Atmospheric Sciences, Kyungpook National University, Daegu 41566, Korea}



\begin{abstract}
We present newly discovered dwarf galaxy candidates in deep and wide-field images 
of NGC 1291 obtained with the Korea Microlensing Telescope Network. We identify 15 dwarf galaxy 
candidates by visual inspection. Using imaging simulations, we demonstrate 
that the completeness rate of our detection is greater than 70\% for the central surface 
brightness value of 
$\mu_{0,R} \lesssim$ 26 \sbr\ and for magnitudes $M_R\lesssim-10$ mag. 
The structural and photometric properties of the dwarf 
galaxy candidates appear to be broadly consistent with those of ordinary 
dwarf galaxies in nearby groups and clusters, with $\mu_{0,R} \sim$ 22.5 to 26.5 
\sbr\ and effective radii of 
200 pc to 1 kpc. The dwarf galaxy candidates show a concentration towards NGC 
1291 and tend to be redder the closer they are to the center, possibly 
indicating that they are associated with NGC 1291.
The dwarf candidates presented in this paper appear to be bluer than those in denser environments,
revealing that the quenching of star formation in dwarf galaxies is susceptible 
to the environment, while the morphology shaping is not.
\end{abstract}


\section{Introduction} \label{sec:intro}
The $\Lambda$CDM paradigm successfully predicts the large-scale structures of 
the Universe, but it often fails to explain the properties on much smaller
scales (see \citealp{2017ARA&A..55..343B}). Together with the too-big-to-fail 
\citep{2011MNRAS.415L..40B} and the planes of satellites problems 
\citep{2012MNRAS.423.1109P,2017A&A...602A.119M}, 
one of the critical issues with this paradigm is the so-called `missing satellites 
problem' (\citealp{1999ApJ...522...82K,1999ApJ...524L..19M}) in which models predict that 
there are more satellite galaxies for a given host galaxy than are actually observed, 
based on the census of dwarf-scale satellite galaxies in the Local Group (LG).

There have been a lot of effort towards explaining the problems using cosmological simulations 
with higher resolution and much improved baryonic physics (see Section 3 in 
\citealp{2017ARA&A..55..343B}). Focusing on the possibility of incomplete observations, 
there have been several scenarios to explain the discrepancy between the predicted and observed number 
of satellite galaxies. One scenario suggests that tidal disruption of 
satellite galaxies can reduce the number of observable dwarf galaxies 
(e.g., \citealp{2004ApJ...609..482K,2017MNRAS.471.1709G}), but leaves behind distinct 
structures such as shells, streams, and tidal tails.
It is also possible to mitigate the problem by applying physical mechanisms 
which suppress star formation in dwarf galaxies, such as 
supernovae-driven outflows (e.g., \citealp{2011MNRAS.417.1260F}). 
Those processes leave remnants that would be expected to have a low surface 
brightness (LSB). In that respect, deep imaging surveys of LSB features are 
important as probes into the evolution of dwarf-scale satellite galaxies.

Since the 1980's when LSB galaxies were first discovered \citep{1984AJ.....89..919S}, 
many studies have reported discoveries of those galaxies despite the surface brightness 
limit ($\mu_{0,V}\lesssim 26$ \sbr; e.g., \citealp{1988ApJ...330..634I,1991ApJ...376..404B,1997AJ....114..635D}). 
Although intensive all-sky surveys, such as the Sloan Digital Sky Survey 
(SDSS; \citealp{2000AJ....120.1579Y}), have made a tremendous contribution to this field, the detection 
of faint dwarf galaxies is still difficult because of their low surface brightness, which have a value less than  
1\% of the brightness of the night sky. 

\begin{deluxetable}{lcc}[t]
\tablecaption{Basic information of NGC 1291 \label{tab:mathmode}}
\tablecolumns{2}
\tablenum{1}
\tablewidth{0pt}
\tablehead{
\colhead{Parameter} &
\colhead{Value} &
\colhead{Reference}
}
\startdata
R.A. (J2000) & $03^h17^m18^s.6$ & 1 \\
Decl. (J2000) & $-41^\circ06^\prime29^{\prime\prime}$ & 1\\
Morphology & (R)SB0/a(s)&1\\
Distance & 9.08 Mpc & 2\\
M$_\star$ & $\sim$$8\times10^{10}$\solmass & 3\\
$m_R$ & 7.98 mag & 4\\
$r_e$ & $\sim$4--5 kpc & 4
\enddata
\vspace{-2mm}
\flushleft
References: (1) NASA Extragalactic Data base (NED); (2) \citet{2017AJ....154...51M}; (3) 
\citet{2011ApJ...737...41L}; (4) measured by \citet{2018AJ....156..249B}.
\end{deluxetable}

Recently, new observational strategies and data analysis techniques for deep imaging surveys 
have been progressively developed to explore the LSB nature in galaxy groups, clusters 
(e.g., \citealp{2011MNRAS.412.1881K,2012ApJS..200....4F,
2015A&A...581A..10C,2015ApJ...798L..45V,2015ApJ...809L..21M,2015ApJ...813L..15M,
2016ApJ...833..168M,2017A&A...608A.142V,2017ApJ...848...19P,2018MNRAS.481.1759K,2018ApJ...863..152S,
2019ApJ...885...88P}), and fields (e.g., \citealt{2016A&A...588A..89J,2017A&A...603A..18H,
2017ApJ...842..133L}). Furthermore, automated detection methods for identifying faint dwarf galaxies have 
been contributed to much larger explorations (e.g., \citealp{2014ApJ...787L..37M,
2016MNRAS.456.1607D,2016MNRAS.456.1359F,2016A&A...590A..20V,2016ApJS..225...11Y,
2017ApJ...850..109B,2018ApJ...857..104G,
2018MNRAS.478..667P,2019ApJS..240....1Z,2019arXiv190907389C}). 
All of these efforts contribute to the detection of LSB galaxy populations and understand 
the evolution of dwarf galaxies in different environments.

In this paper, we present a list of dwarf galaxy candidates which are newly discovered in deep optical 
images of NGC 1291 taken by the Korea Microlensing Telescope Network 
(KMTNet; \citealt{2016JKAS...49...37K}). NGC 1291 is located at a distance of 9.08 Mpc 
\citep{2017AJ....154...51M} and there are two potential companion galaxies that have similar 
recessional velocities (see Section 4.2). The survey of the NGC 1291 system, which has been thought 
to be a low-density environment, can contribute to the study of environmental effects on the evolution 
of dwarf galaxies. The basic properties of NGC 1291 are presented in Table 1. 
The KMTNet imager contains four CCD chips with a field-of-view (FoV) of $\sim$1 deg$^2$ each, so 
the final mosaic images cover the whole area within the virial radius of 
NGC 1291\footnote{To obtain the virial radius of NGC 1291, we 
adopted the conversion equation of $r_{1/2}\sim 0.015r_{200}$ \citep{2013ApJ...764L..31K}. 
We assumed that the half-mass radius ($r_{1/2}$) can be replaced with the effective radius ($r_e$) 
because of a flat color profile in NGC 1291 \citep{2018AJ....156..249B}.}. 
In this study, we assumed that all of the dwarf galaxy candidates are located at the same 
distance as NGC 1291.

This paper is organized as follows. Section 2 describes the outline of the observations and the data 
reduction process. Section 3 presents a list of dwarf galaxy candidates, the completeness of 
detections, and the properties of the discovered dwarf galaxy candidates. Section 4 discusses the 
implications of the results for the evolution of dwarf galaxy candidates. Section 5 summarizes the 
results. 

\begin{figure*}[t]
\centering
\includegraphics[width=85mm]{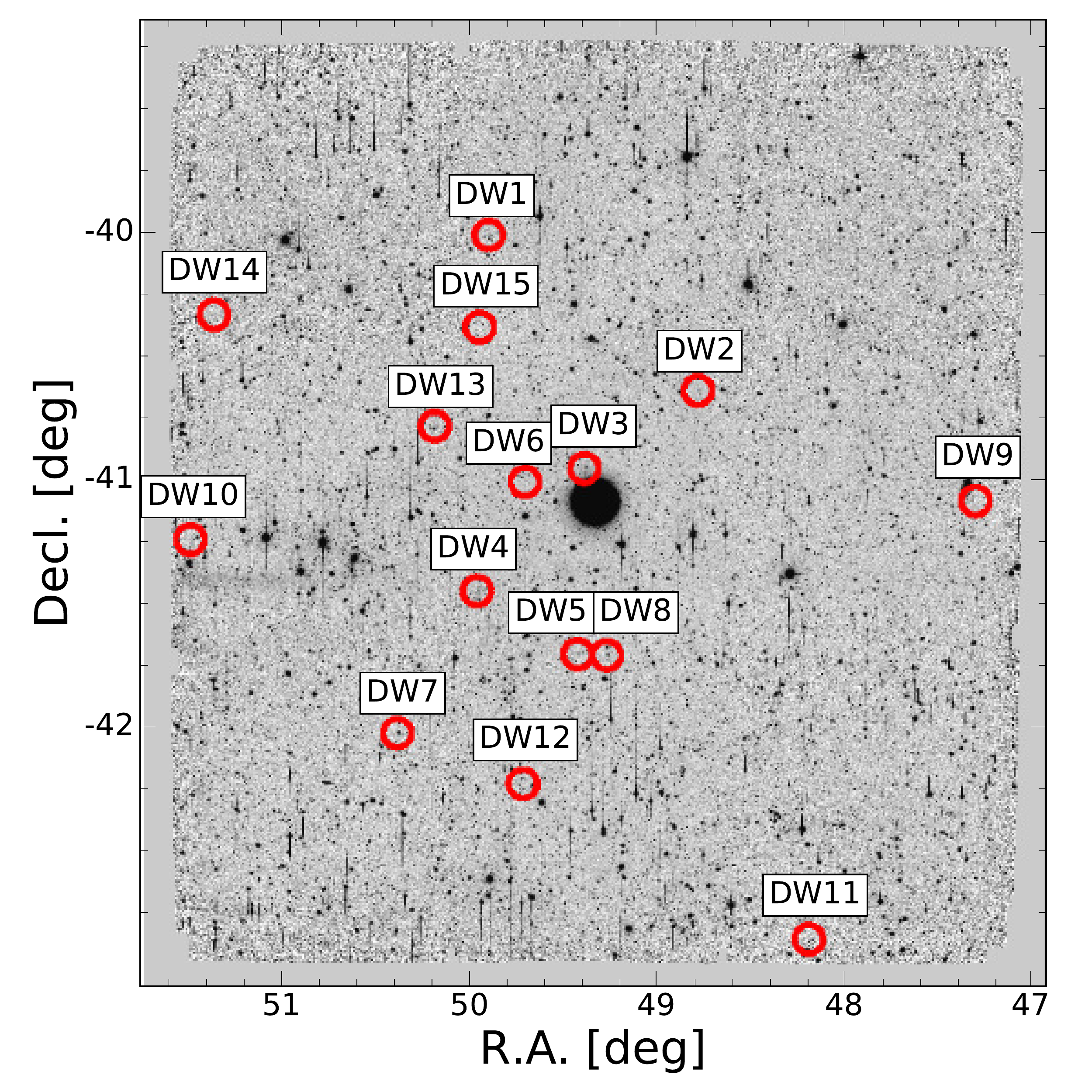}
\includegraphics[width=85mm]{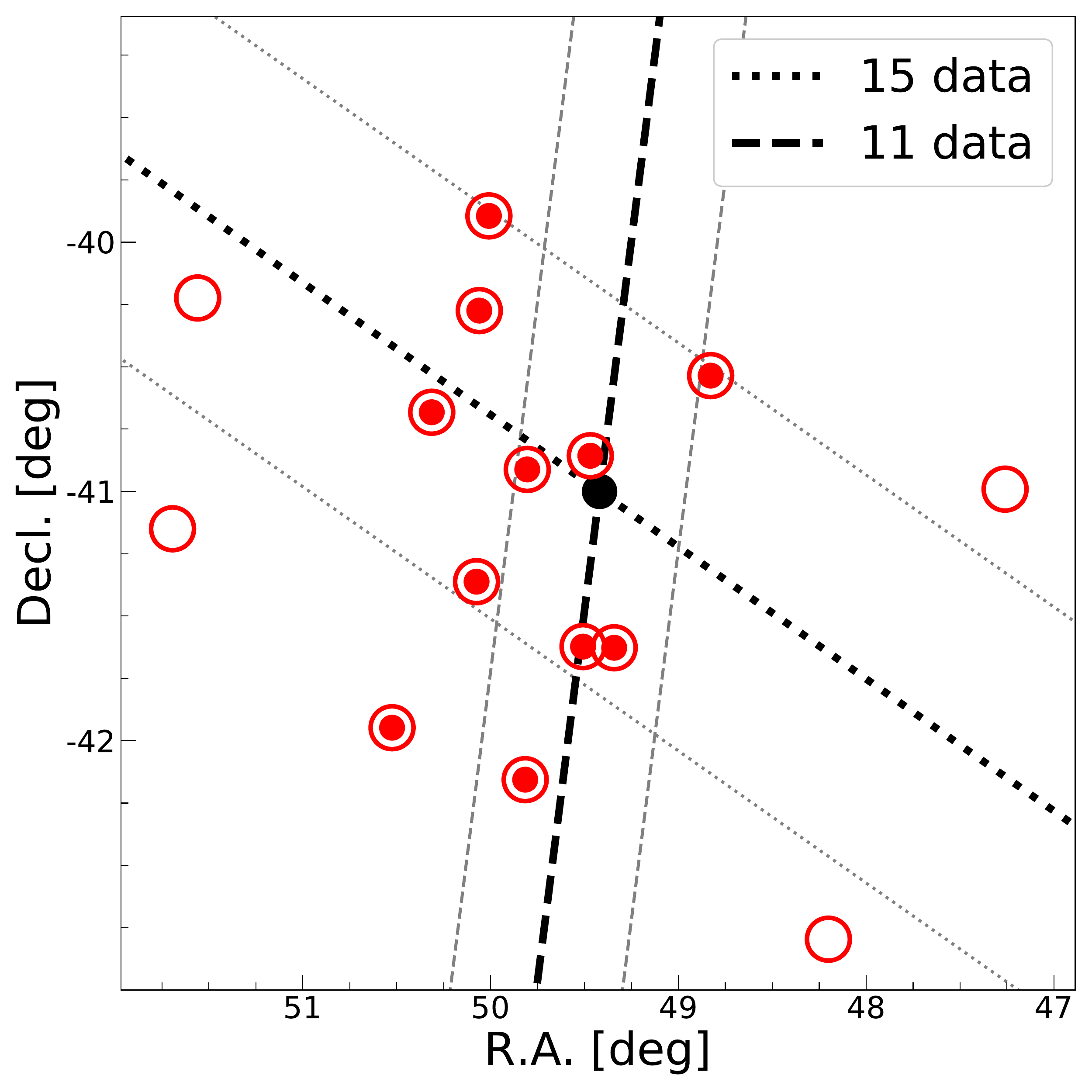}
\caption{Left: The spatial distributions of 15 dwarf galaxy candidates (red circles) that we identified in 
the $R$-band mosaic image. NGC 1291 is located at the center. One side of the image is equivalent to 
$\sim$600 kpc at the distance of NGC 1291. North is up and east is to the left. Right: The potential 
planes of satellite galaxies with the whole data (open circles; dotted line) and those relatively close to 
NGC 1291 (semi-filled circles; dashed line). The gray lines represent $\pm1\sigma$.} 
\label{fig:fig1}\vspace{5mm}
\end{figure*}

\section{Data}
We conducted deep optical imaging of NGC 1291 with KMTNet. 
Consisting of three 1.6-m identical telescopes, KMTNet is located at the Cerro Tololo Inter-American 
Observatory (CTIO), the South African Astronomical Observatory (SAAO), and Siding Spring 
Observatory (SSO). Each telescope contains a wide-field CCD camera with a $2^{\circ}\times2^{\circ}$ FoV
with average pixel scale of 0.4 arcsec. The data were taken at the 
KMTNet-CTIO observatory on November 12, 2015 with an average seeing of 
$\sim$1.1 arcsec. 
A total of 85 $B$-band and 48 $R$-band images were utilized which had total exposure times of 
$\sim$2.8 and $\sim$1.6 hrs, respectively. 
The images were taken by placing the 
target galaxy on each of the four chips with additional 7-points dithers. This strategy enabled us to obtain 
blank sky regions for dark-sky flat-fielding and also fill in the CCD gaps in the stacked images.
The overall description of KMTNet Nearby Galaxy Survey (KNGS) project is also found in 
\citet{2018AJ....156..249B}.

The data reduction procedure used in this study is the same as that used in the previous 
study except for the addition of crosstalk correction \citep{2016PKAS...31...35K}. 
We utilized crosstalk-corrected data this time to avoid 
confusing faint dwarf galaxies and artificial features. Brief description of the data reduction 
is as follows.
\begin{itemize}
\item {\it Overscan correction} was adopted to subtract bias level because of the apparent instability 
of the bias frames.
\item {\it Dark sky flat} was generated by stacking all object frames with object-mask and then applied for 
flat-fielding.
\item {\it A two-dimensional sky model} was subtracted from each object frame to remove gradients over 
the sky background.
\item Accurate astrometric calibration was performed with {\tt SCAMP} \citep{2006ASPC..351..112B} 
following the instructions provided by the KMTNet project team\footnote{\url{http://kmtnet.kasi.re.kr}, 
``Astrometric calibration for KMTNet data''}. 
\item Stacked images were generated by {\tt SWarp} \citep{2002ASPC..281..228B} using 
{\it median-combine} without photometric calibration.
\end{itemize}

Finally, we obtained deep mosaic images with a FoV of 12 deg$^2$ 
centered on NGC 1291 in the $B$- and $R$-bands. The photometric zero points for the mosaic images 
were determined using the AAVSO Photometric All-Sky Survey (APASS) DR9 
catalog\footnote{\url{https://www.aavso.org/apass}, We derived $R$ magnitude using the conversion 
equation $R=r-0.1837\times(g-r)-0.0971$ provided by \citet{lupton05}}. 
The optimized data reduction procedure allowed to investigate the surface brightness down to 
$\sim$28--29 \sbr\ within 1$\sigma$ limit. 

\section{Results}
\subsection{Detection}
Four of us (W. Byun, Y.-K. Sheen, H. S. Park, and M. Kim) performed visual inspection for 
searching faint dwarf galaxies, independently. The dwarf galaxy candidates had to 
(1) have similar appearances in both $B$- and $R$-band images; 
(2) exhibit a smooth gradation of surface brightness on their outskirts to distinguish them from 
massive background galaxies; and (3) not be located at the vicinity of bright stars that may 
interfere with accurate identification. Each object discovered by each of us was then 
independently graded as ``A/B/C'', where ``A" is assigned to strong candidates with low surface 
brightness and extended light, ``B" is assigned to ambiguous candidates with very low surface 
brightness, that can be easily confused with artifacts (e.g., crosstalk), ``C" is assigned to less 
promising candidates with compact features. While a total of 35 
objects were initially identified, we assessed that it might be biased by human subjectivity. 
Hence, we utilized the objects which are detected by at least two authors regardless of their 
grades. Then, we converted the grades ``A/B/C" into the scores 25/15/5, and the 
objects with cumulative scores higher than 65 were designated as dwarf galaxy candidates. 
For the ambiguous candidates, several authors closely examined the images together for 
the final decision. This process resulted in the removal of probable background objects and tidal features 
from the list. A total of 15 dwarf galaxy candidates were identified and their locations are shown in Figure 1. 
Four out of our candidates (N1291-DW6/DW7/DW8/DW14) had been previously detected 
\citep{1999MNRAS.304..297M,2003A&A...412...45P}, but the rest are newly discovered in this study. 

\begin{figure}[t]
\centering
\includegraphics[width=85mm]{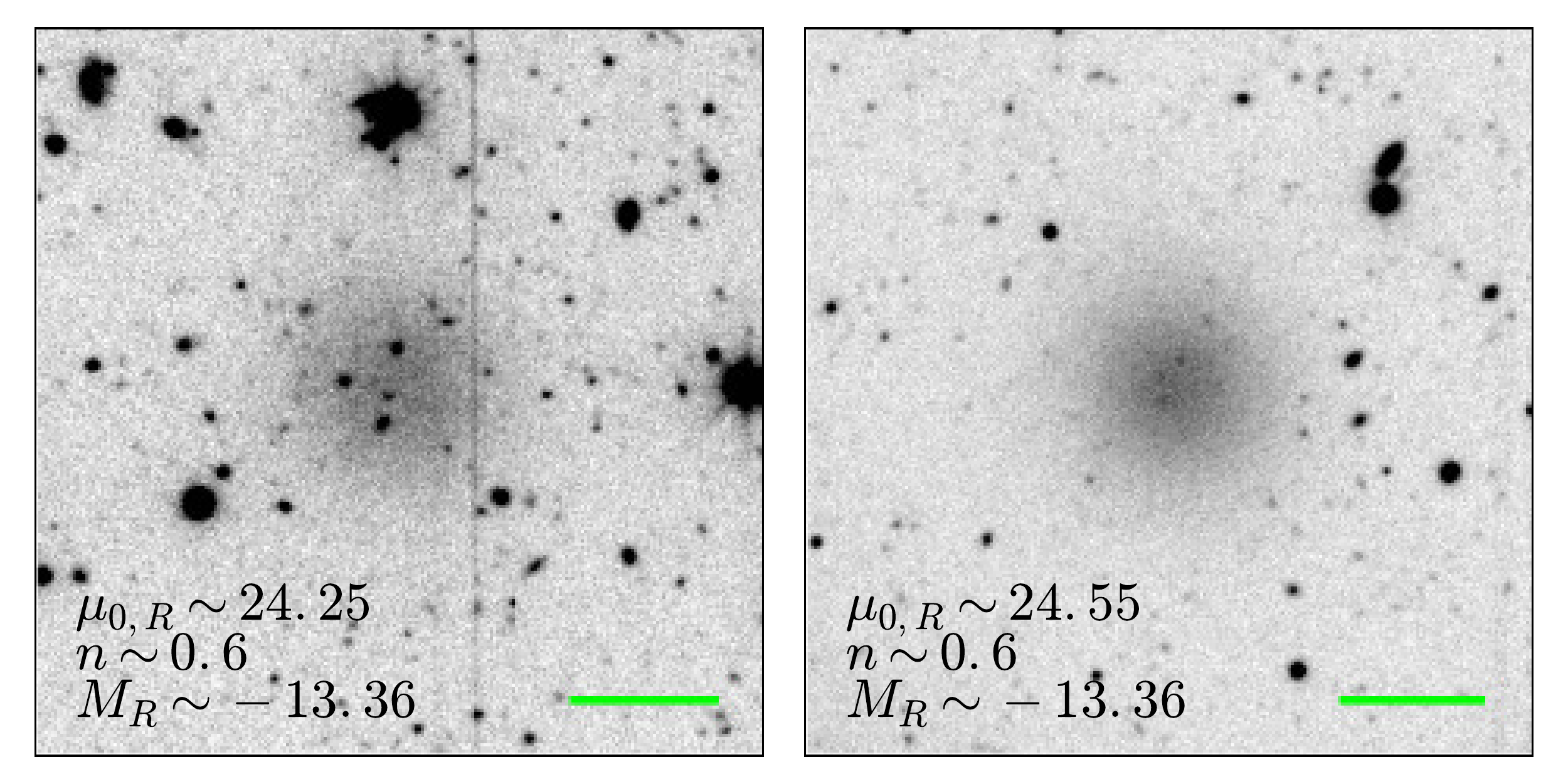}
\caption{Comparison between the dwarf candidate N1291-DW4 (left) and a simulated 
dwarf galaxy (right). Both dwarfs share similar structural and photometric 
properties. The central surface brightnesses, S\'ersic indices and absolute 
magnitudes are shown in the images. The horizontal bar at the bottom-right 
corresponds to 30 arcsec.
} \label{fig:fig2}
\end{figure}

\begin{figure*}[t]
\centering
\includegraphics[height=85mm]{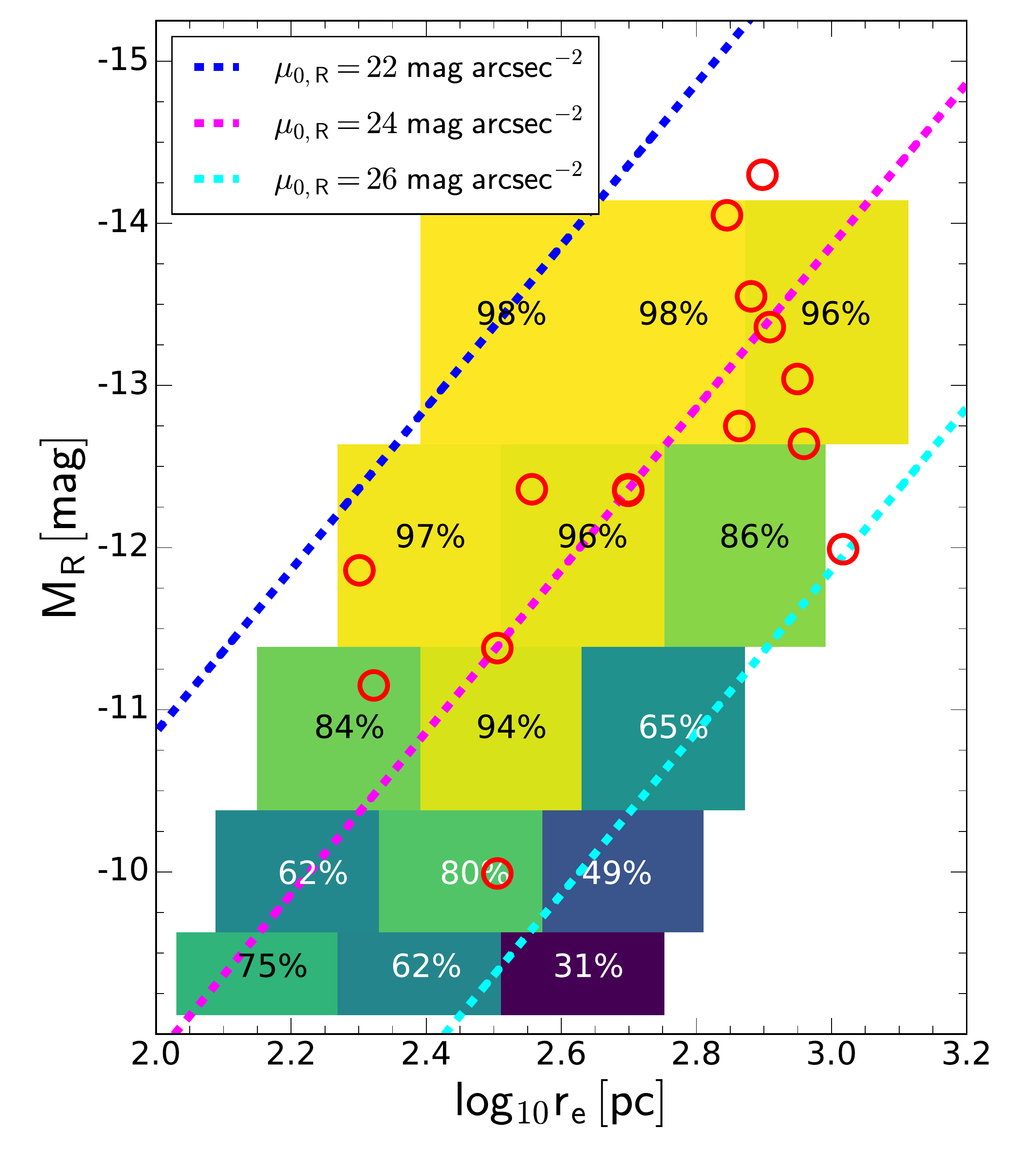}
\includegraphics[height=85mm]{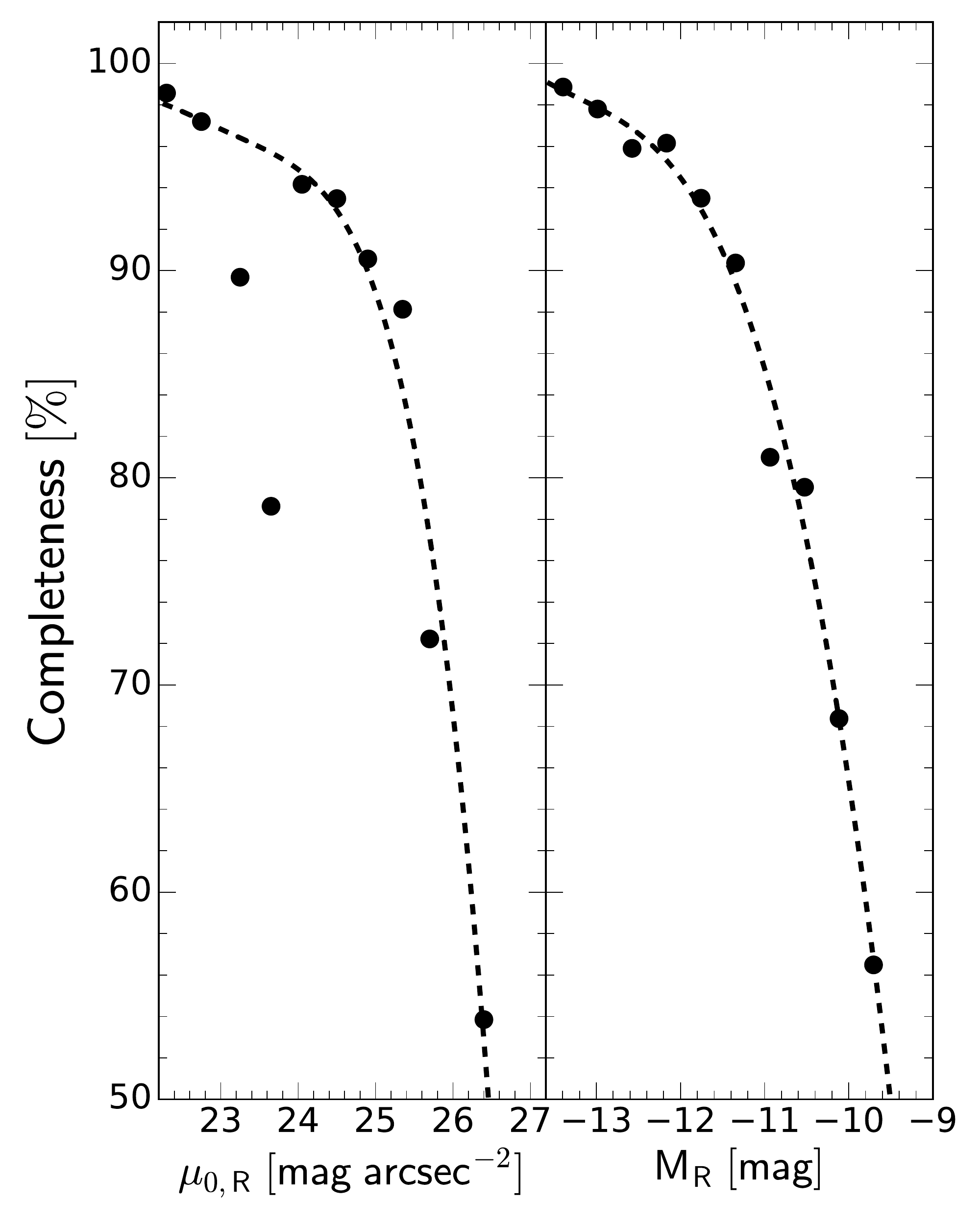}
\caption{Left: Recovered fractions for simulated galaxies in absolute 
magnitude-effective radius space. The completeness rate of each bin is color-coded and noted.
The red circles represent the dwarf galaxy candidates discovered in the NGC 1291 system. 
The dashed lines represent the central surface brightnesses 
of 22 (blue), 24 (magenta), and 26 (cyan) \sbr\ when $n=0.75$.
Right: Completeness rates as a function of central surface brightness and absolute magnitude. 
The dashed lines are derived by quartic polynomial fit. 
} \label{fig:fig3}
\end{figure*}

Interestingly, most of the dwarf candidates were located on the eastern side of NGC 1291 and were 
distributed from north to south. 
Indeed, such structures have been reported in other studies (e.g., \citealt{2007MNRAS.374.1125M,
2013Natur.493...62I,2014ApJ...787L..37M,2015ApJ...802L..25T,2017A&A...602A.119M}). 
If the satellites are isotropically distributed around the host galaxy, 
there must be no dependency for the specific direction. So we performed a total least 
square fit using the dwarf candidates to find potential plane structures. As seen in the right panel of 
Figure 1, we divided the sample into two subsamples taking into account the uncertainty of virial radius of NGC 
1291: $D_{proj}\lesssim300$ kpc (open circles) and $D_{proj}\lesssim250$ kpc (semi-filled circles). 
When using the whole data set, the dwarf candidates seemed to be aligning with a large scatter from 
the northwest to the southwest (dotted lines). Meanwhile, when only 11 candidates were used, they seemed to 
be more tightly aligned along the north-south direction (dashed lines). 
Note that such analyses can be affected by a projection effect, 
so further spectroscopic observations may be needed to confirm the planarity of the satellite galaxies. 

\subsection{Imaging simulation}
Imaging simulations were conducted to estimate the completeness of our visual inspection. 
Mock galaxies were generated 
on top of the mosaic image using {\tt photutils} and {\tt astropy.modeling} 
library in the PYTHON package. The mock galaxies were modeled by adopting 
the parameters of known dwarfs; 
(1) $-9.5 \le M_R \le -13.5$ mag, (2) $0.1 \le r_e \le 1$ kpc, (3) $0.5\le$ S\'ersic index $n \le 1.0$, and (4) 
ellipticity $e$ $\le 0.6$. These mock galaxies were distributed randomly in the image and 
the line-of-sight distances were set to follow a Lorentzian distribution with a 
variance of $\pm$1 Mpc\footnote{To reflect the actual distribution of dwarf 
galaxies in MW-like galaxies, we employ an additional constraint that 
$\sim$10\% of simulated galaxies have $\Delta d>\pm0.4$-1 Mpc.} (cf. \citealt{2018ApJ...863..152S}). 
The large wings in Lorentzian distribution can affect the completeness, but its effect appeared to 
be negligible because most of the simulated galaxies were located within a line-of-sight distance of 
$\sim$ 300--400 kpc. Figure 2 compares a simulated galaxy with a dwarf galaxy candidate discovered in this study. 

\subsubsection{Completeness}
Eight imaging simulations were performed. In each simulation, $\sim$150 mock galaxies were generated. 
In order to estimate the completeness rate, the mock images were again subject to visual inspection. 
Note that we did not control the mock galaxies which are overlapped with bright objects, 
so the following results are presented as the lower limit. As shown in the left panel of Figure 3, the completeness 
rate reaches over $\sim$90\% for $M_R\lesssim-11.5$ mag in all galaxy sizes and 
$\sim$70\% for $M_R\lesssim-10$ mag.
Meanwhile, the completeness rate appears to be much sensitive to the galaxy size or central surface brightness
 for $M_R\gtrsim-11$ mag. 
This result is mainly due to confusion between small dwarf galaxies and massive background 
galaxies, and also partly induced by the 3$\sigma$ surface brightness limit of $\sim$26.5 \sbr\ in the mosaic image (see 
\citealp{2018AJ....156..249B}). The right panel of Figure 3 also shows the average completeness rates against 
central surface brightness and absolute magnitude. Although the points are sparsely distributed, the fitted curves 
support that the completeness rate reaches over $\sim$70\% for $M_R\lesssim-10$ mag and $\mu_{0,R}\lesssim$ 26 \sbr. 

\begin{deluxetable*}{lcccccccccc}[t]
\tabletypesize{\scriptsize}
\tablecaption{Catalog of the Dwarf Galaxy Candidates in the NGC 1291 system\label{tab:mathmode}}
\tablecolumns{6}
\tablenum{2}
\tablewidth{0pt}
\tablehead{
\colhead{ID} & \colhead{R.A.(J2000)} & \colhead{Decl.(J2000)} &
\colhead{$m_R$\tablenotemark{a}} & \colhead{$B-R$\tablenotemark{b}} &
\colhead{$\mu_{0,R}$\tablenotemark{c}} &\colhead{$n$\tablenotemark{c}} & 
\colhead{$r_e$\tablenotemark{c}} & \colhead{$M_R$\tablenotemark{d}} & 
\colhead{$\log$ $M_\star$\tablenotemark{e}}\\
\colhead{} & \colhead{(hh:mm:ss)} & \colhead{(dd:mm:ss)} &
\colhead{(mag)} & \colhead{(mag)} & \colhead{(mag arcsec$^{-2}$)} &
\colhead{} & \colhead{(\arcsec)} & \colhead{(mag)}&
\colhead{(\solmass)} & \colhead{}
}
\startdata
N1291-DW1 & 3:19:25.9 & -40:03:47 & 17.46$\pm$0.13 & 1.31 & 24.18$\pm$0.19 &  0.65$\pm$0.03 & 11.30$\pm$0.23 & -12.34 & 6.97 \\
N1291-DW2 & 3:15:09.7 & -40:40:05 & 17.05$\pm$0.13 & 1.17 & 24.61$\pm$0.19 & 0.65$\pm$0.03 & 16.62$\pm$0.38 & -12.75 & 7.03 \\
N1291-DW3 & 3:17:29.4 & -40:58:22 & 16.25$\pm$0.05 & 1.36 & 23.81$\pm$0.07 & 0.71$\pm$0.01 & 17.30$\pm$0.17 & -13.55 & 7.45 \\
N1291-DW4 & 3:19:43.1 & -41:26:54 & 16.44$\pm$0.08 & 1.28 & 24.25$\pm$0.13 & 0.60$\pm$0.02 & 18.47$\pm$0.26 & -13.36 & 7.42  \\
N1291-DW5 & 3:17:38.1 & -41:41:45 & 16.76$\pm$0.10 & 1.16 & 24.57$\pm$0.14 & 0.77$\pm$0.02 & 20.32$\pm$0.42 & -13.04 & 7.15 \\
N1291-DW6\tablenotemark{f} & 3:18:43.0 & -41:01:27 & 17.44$\pm$0.18 & 1.14 & 23.09$\pm$0.24 & 0.88$\pm$0.04 & 8.24$\pm$0.17 & -12.36 & 6.87 \\
N1291-DW7\tablenotemark{f} & 3:21:24.4 & -41:59:44 & 15.50$\pm$0.04 & 1.13 & 23.34$\pm$0.06 & 0.58$\pm$0.01 & 17.86$\pm$0.14 & -14.30 & 7.48 \\
N1291-DW8\tablenotemark{f} & 3:17:01.4 & -41:42:01 & 15.75$\pm$0.10 & 1.23 & 22.95$\pm$0.13 & 0.82$\pm$0.02 & 15.88$\pm$0.24 & -14.05 & 7.59 \\
N1291-DW9 & 3:09:25.1 & -41:04:57 & 18.42$\pm$0.18 & 1.20 & 23.90$\pm$0.23 & 0.84$\pm$0.04 & 7.37$\pm$0.14 & -11.38 & 6.47 \\
N1291-DW10 & 3:25:38.4 & -41:13:51 & 17.16$\pm$0.23 & 1.11 & 25.08$\pm$0.32 & 0.72$\pm$0.05 & 20.60$\pm$0.82 & -12.64 & 6.99 \\
N1291-DW11 & 3:12:44.2 & -42:47:54 & 17.51$\pm$0.32 & 0.81 & 23.99$\pm$0.44 & 0.80$\pm$0.07 & 11.63$\pm$0.54 & -12.36 & 6.59 \\
N1291-DW12 & 3:18:47.0 & -42:11:58 & 19.81$\pm$0.36 & 1.12 & 25.19$\pm$0.54 & 0.75$\pm$0.09 & 7.22$\pm$0.26 & -9.99 & 5.98 \\
N1291-DW13 & 3:20:33.8 & -40:48:18 & 17.81$\pm$0.34 & 1.22 & 26.39$\pm$0.55 & 0.48$\pm$0.08 & 23.60$\pm$1.89 & -11.99 & 6.62 \\
N1291-DW14\tablenotemark{f} & 3:25:02.8 & -40:21:32 & 17.94$\pm$0.48 & 1.02 & 22.66$\pm$0.63 & 0.80$\pm$0.11 & 4.56$\pm$0.18 & -11.86 & 6.45 \\
N1291-DW15 & 3:19:37.7 & -40:25:17 & 18.65$\pm$0.32 & 1.07 & 22.83$\pm$0.38 & 1.07$\pm$0.22 & 4.82$\pm$0.22 & -11.15 & 6.20 \\
\enddata
\tablecomments{Magnitudes and surface brightnesses are uncorrected for Galactic extinction.}
\tablenotetext{a}{Apparent magnitude calculated by Eq. 2.}
\tablenotetext{b}{Color measured within 2 effective radius derived from $R$-band image.}
\tablenotetext{c}{Central surface brightness ($\mu_{0,R}$), S\'ersic index ($n$) and effective radius ($r_e$) 
in $R$-band image derived from the 1-D profile fit.}
\tablenotetext{d}{Absolute magnitude calculated by assuming that the dwarfs are at the same distance as 
NGC 1291 ($d=9.08$ Mpc).}
\tablenotetext{e}{Stellar mass estimated with an equation of $\log(M_\star/L)=0.872\times(B-R)-0.866$ adopted 
from  \citet{2013MNRAS.430.2715I}.}
\tablenotetext{f}{Cross-ID: [MDS99] F301-056, [MDS99] F301-044, [MDS99] F301-064, LEDA 587327, respectively.}
\end{deluxetable*}

\begin{figure*}[t]
\centering
\includegraphics[width=150mm]{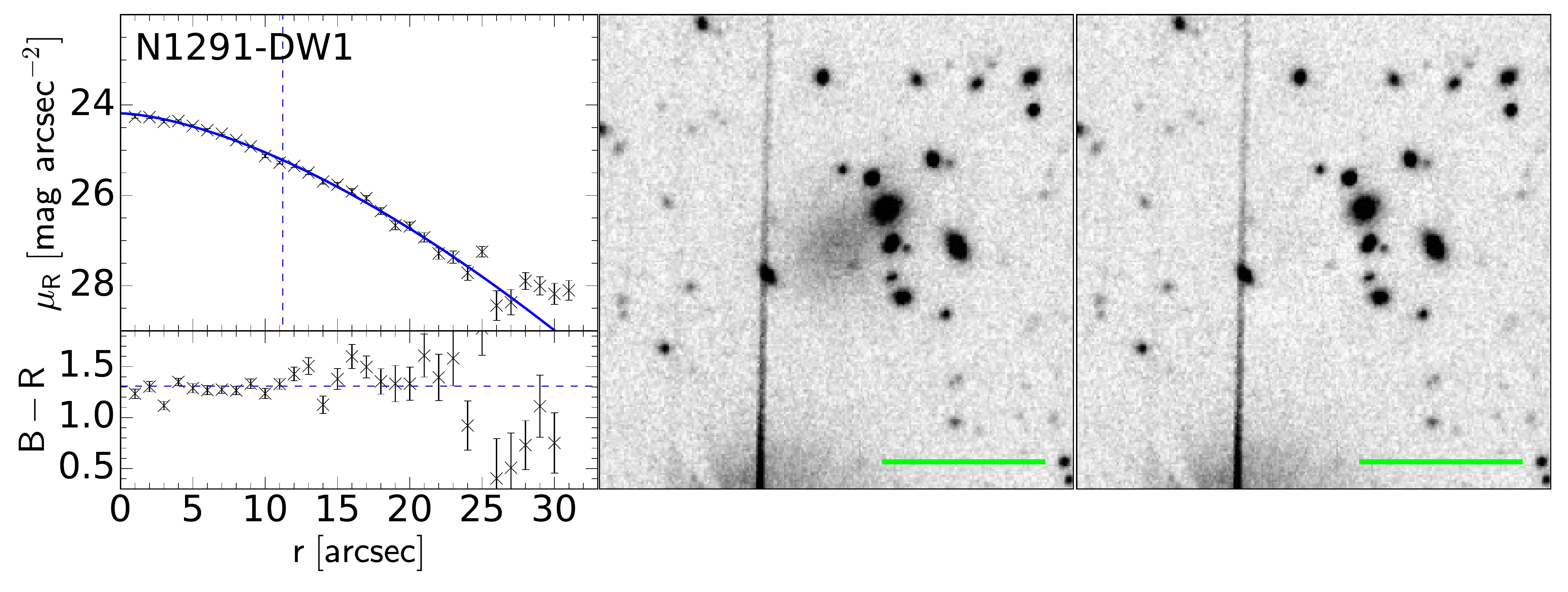}\vspace{-6mm}
\includegraphics[width=150mm]{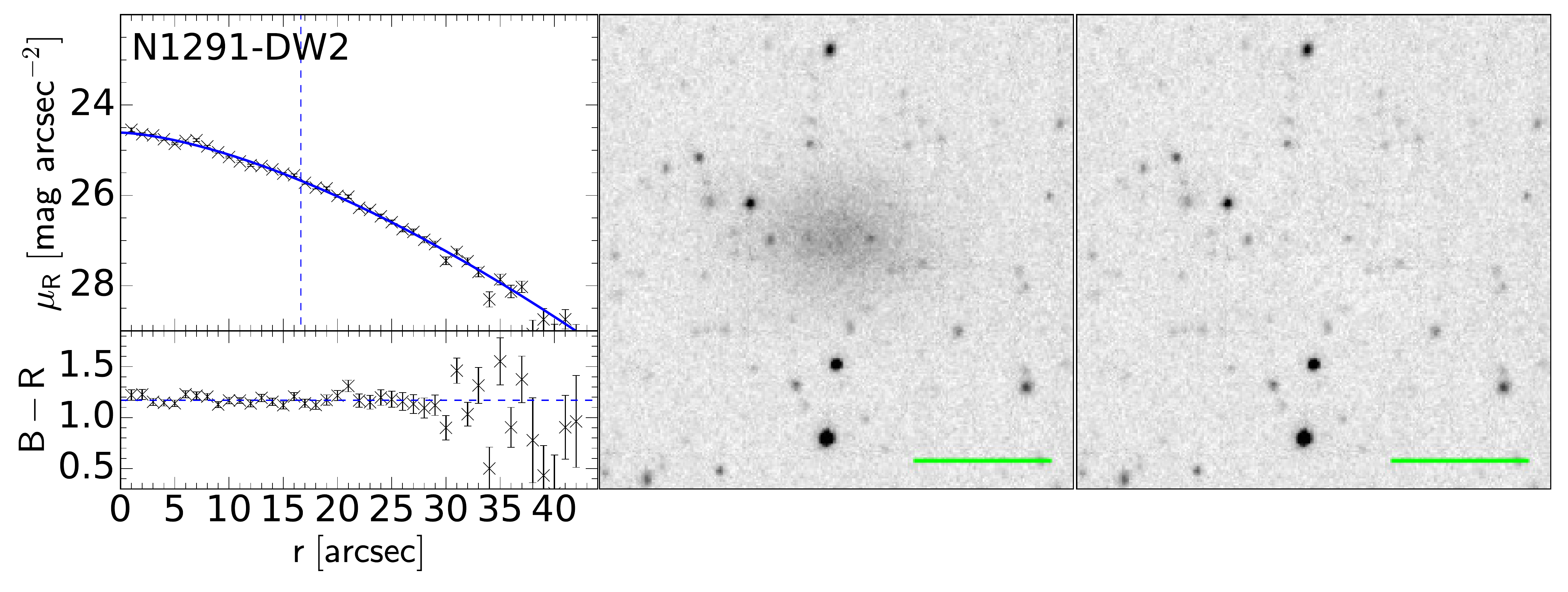}\vspace{-6mm}
\includegraphics[width=150mm]{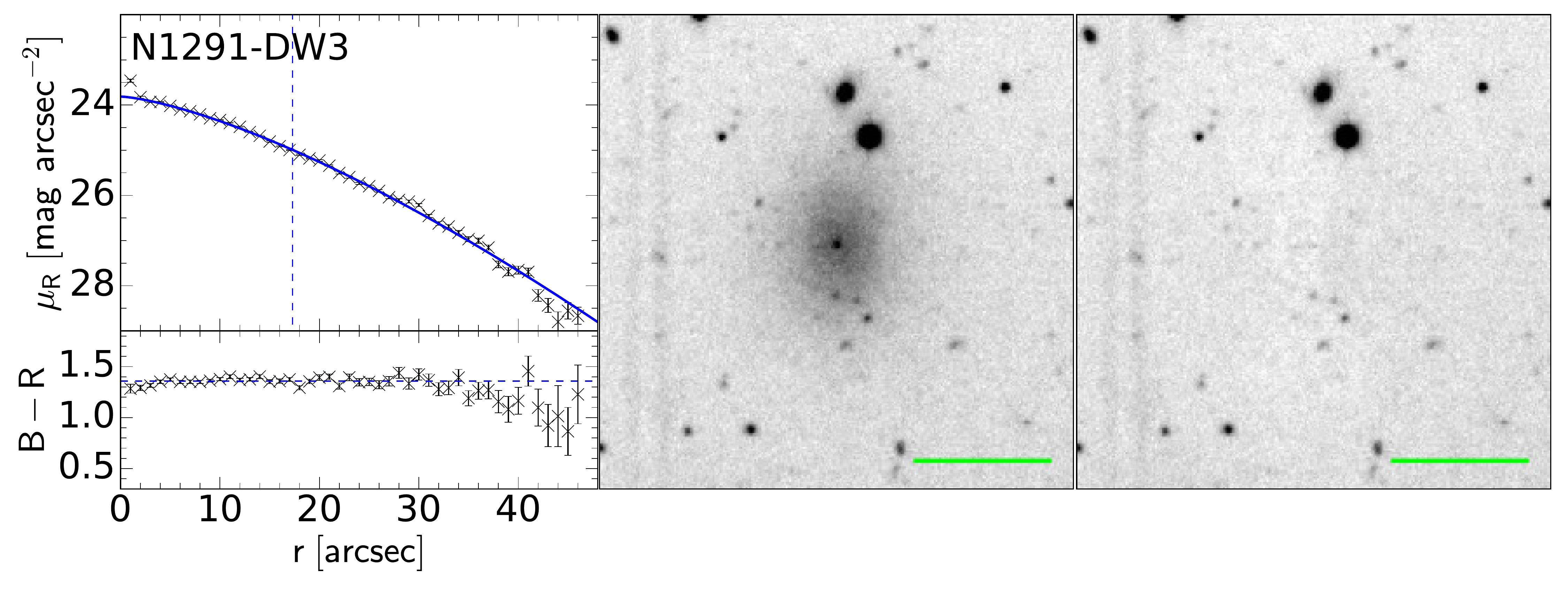}\vspace{-6mm}
\includegraphics[width=150mm]{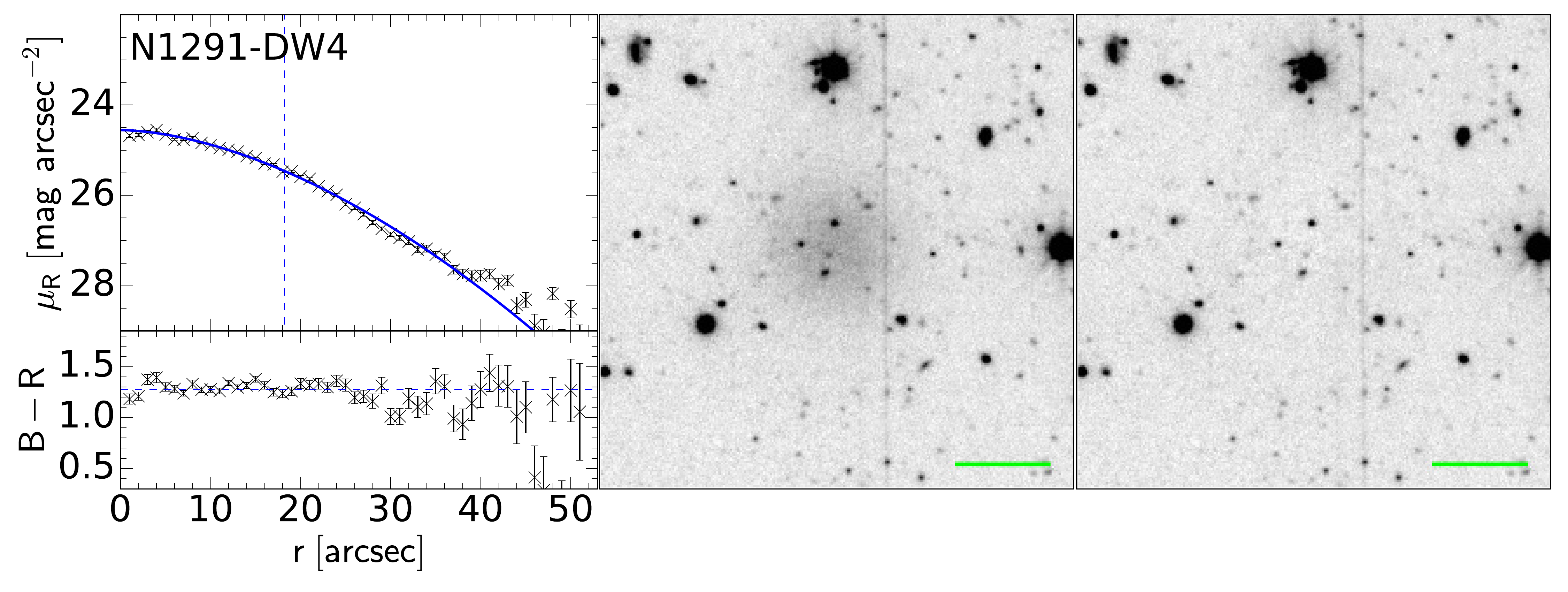}
\caption{The surface brightness and color profiles (left), $R$-band cutout images (middle), 
and residuals (right) of dwarf galaxy candidates as a result of isophotal fit.
The blue solid lines in the left panels represent a single S\'ersic function. The blue dashed 
lines show the effective radii (vertical) and weighted mean colors (horizontal), respectively.
The scale bar in the cutout images corresponds to 30 arcsec.
} \label{fig:fig4}
\end{figure*}
\setcounter{figure}{3}

\begin{figure*}[]
\centering
\includegraphics[width=150mm]{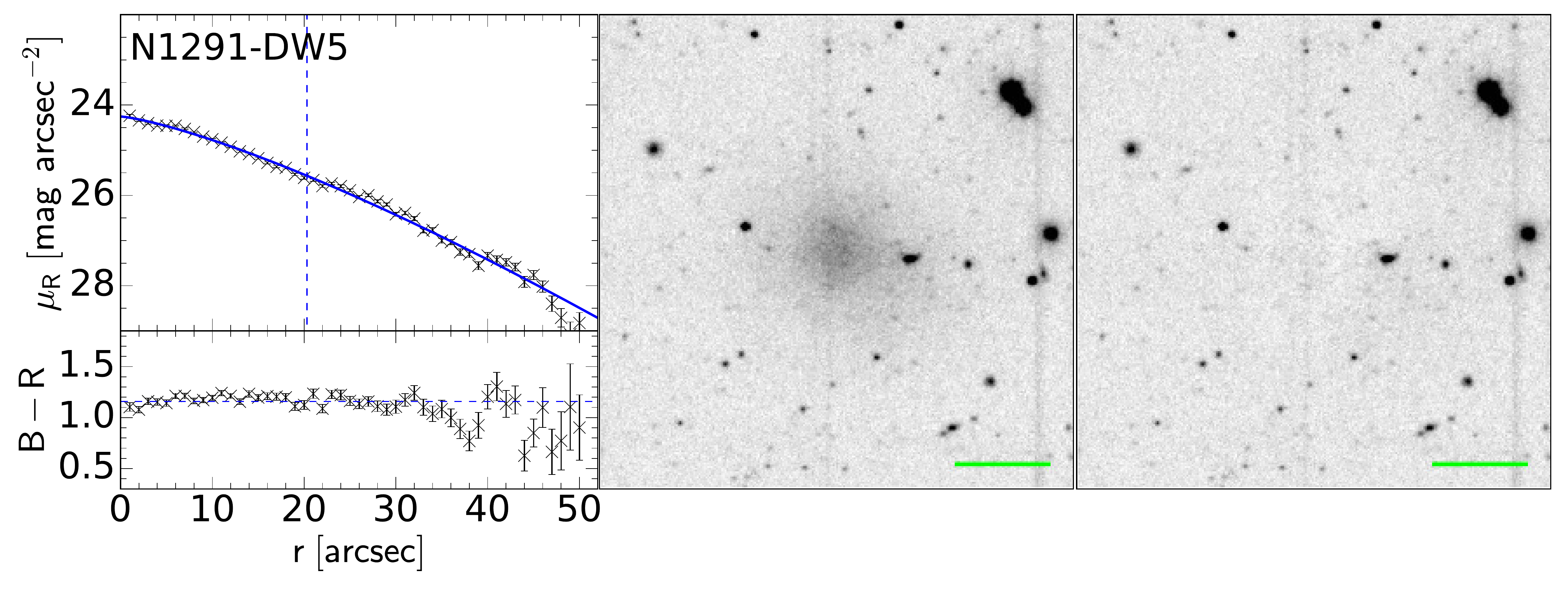}\vspace{-6mm}
\includegraphics[width=150mm]{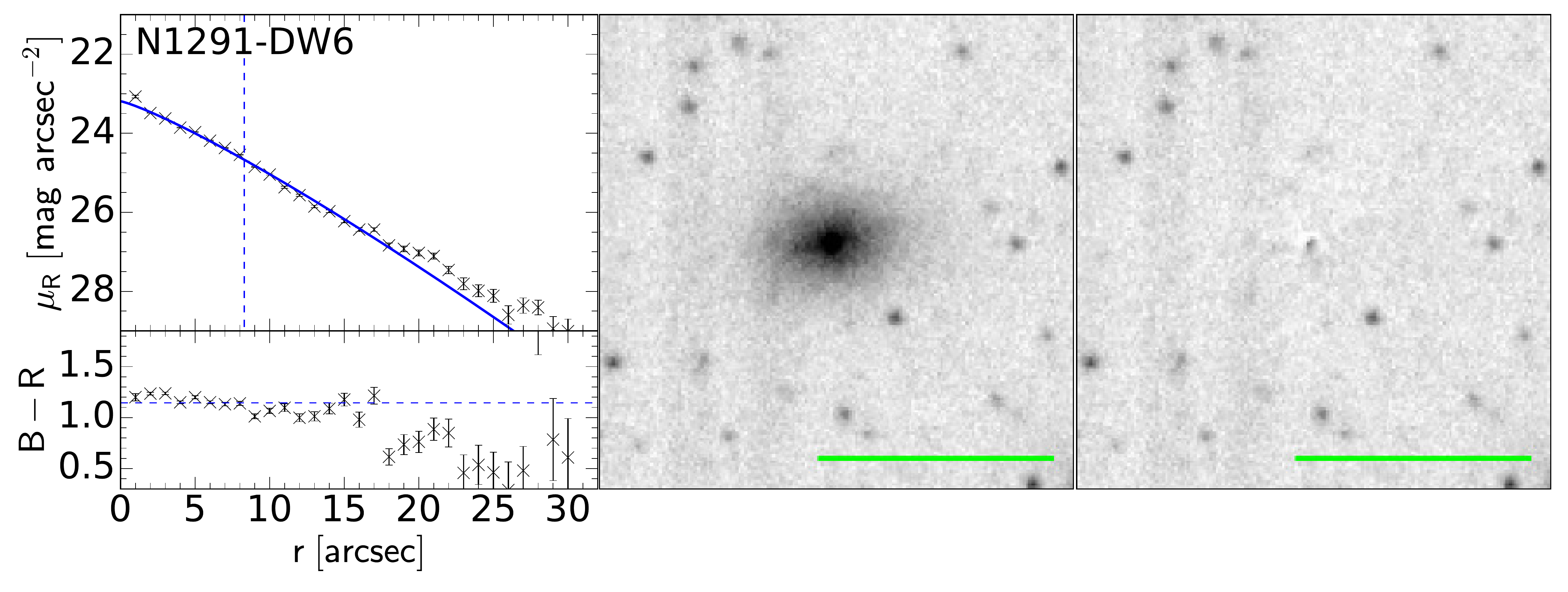}\vspace{-6mm}
\includegraphics[width=150mm]{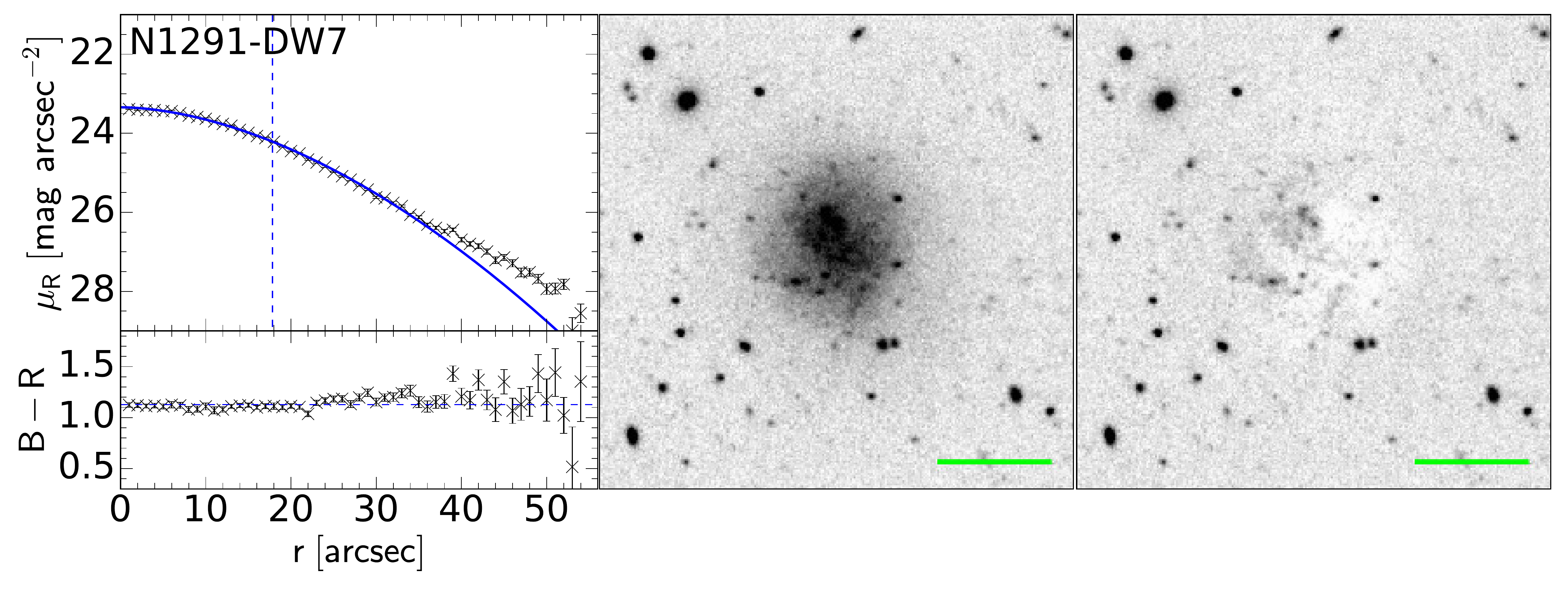}\vspace{-6mm}
\includegraphics[width=150mm]{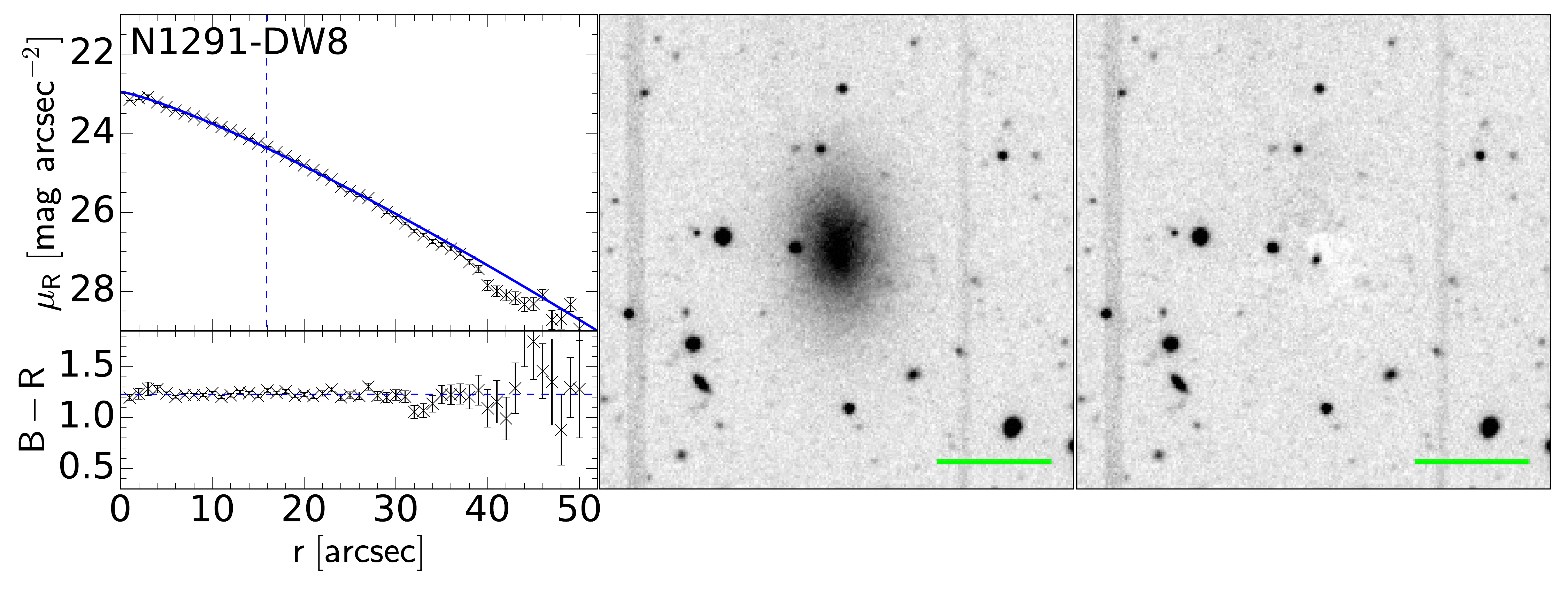}
\caption{(Continued)
} \label{fig:fig4}
\end{figure*}
\setcounter{figure}{3}

\begin{figure*}[]
\centering
\includegraphics[width=150mm]{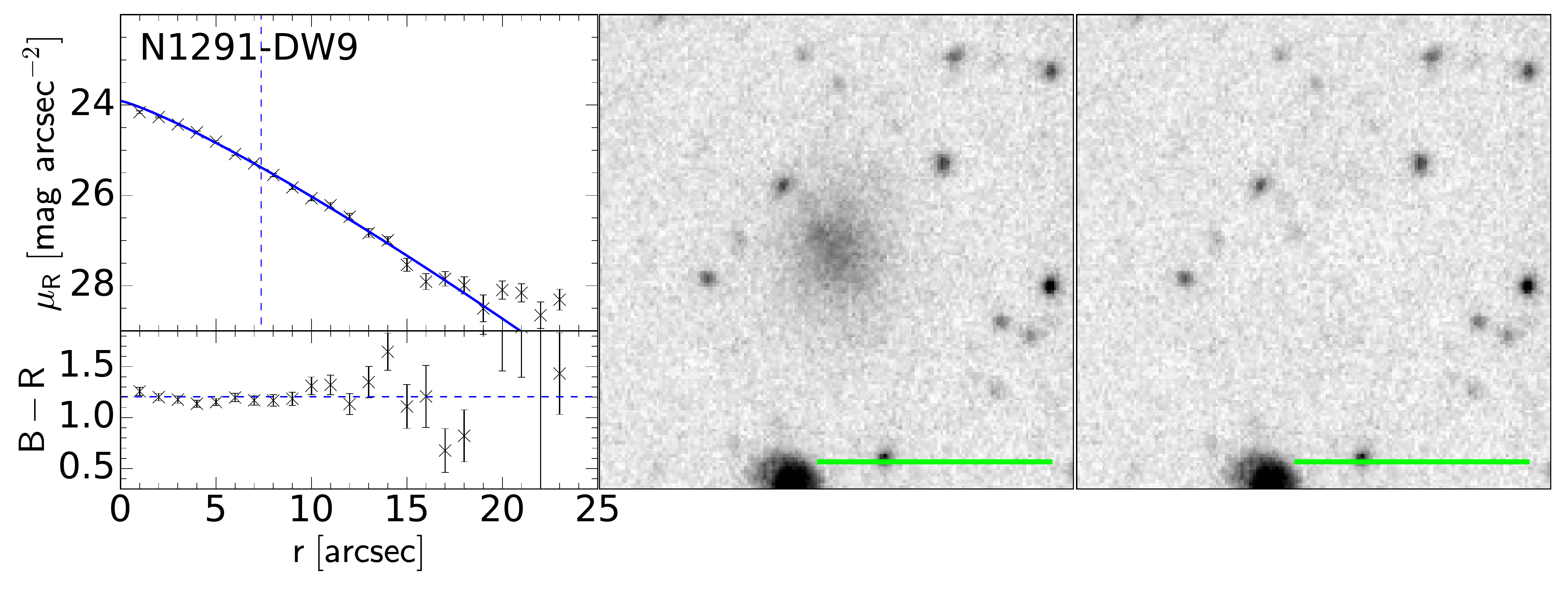}\vspace{-6mm}
\includegraphics[width=150mm]{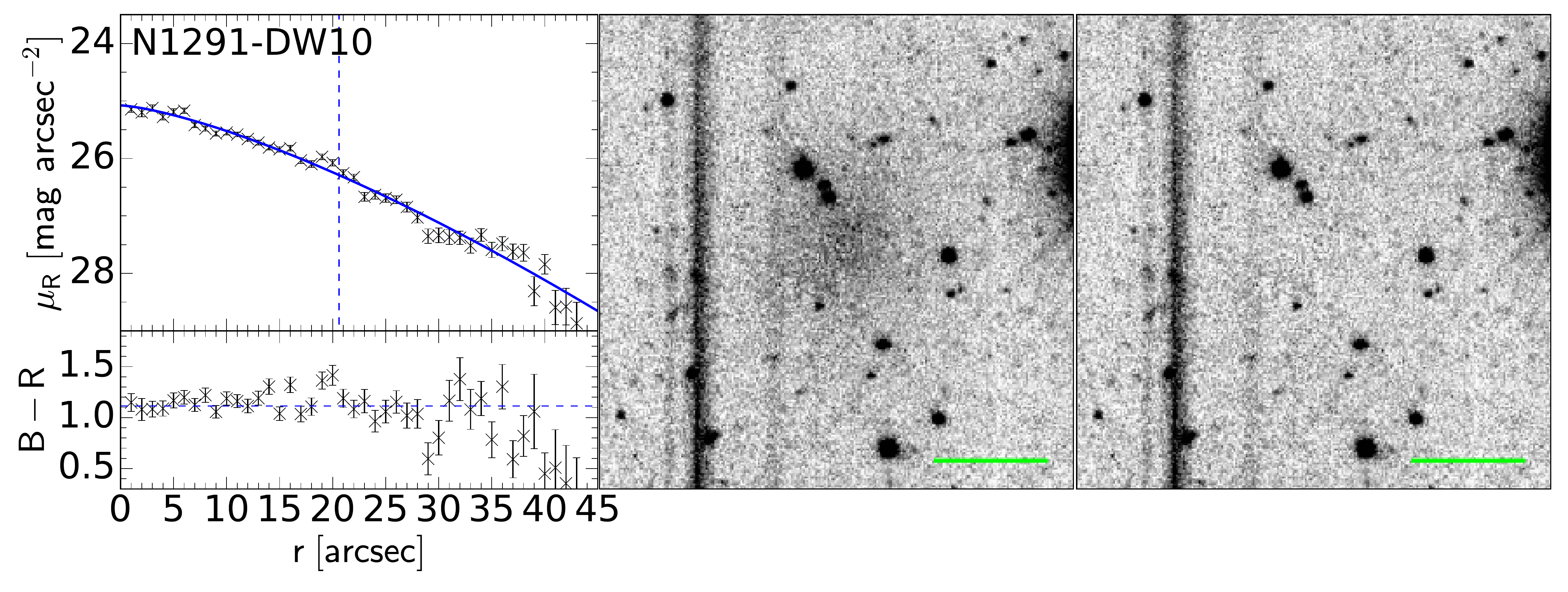}\vspace{-6mm}
\includegraphics[width=150mm]{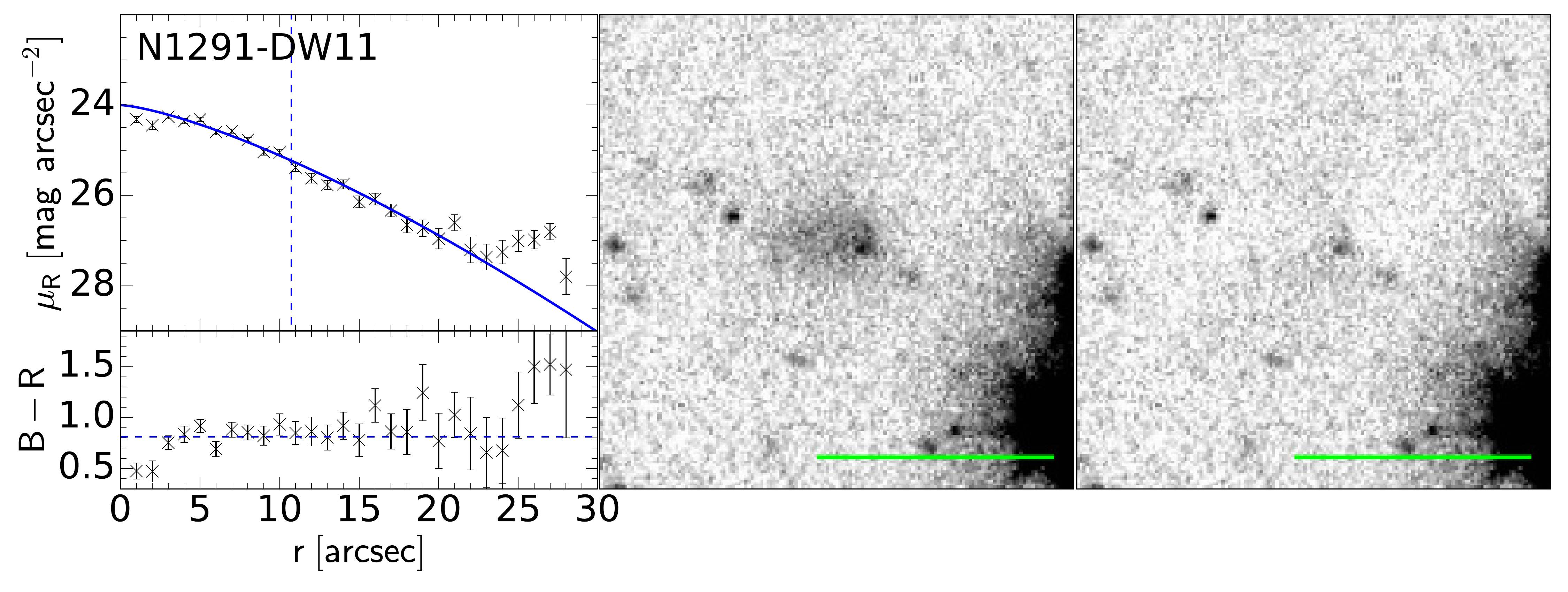}\vspace{-6mm}
\includegraphics[width=150mm]{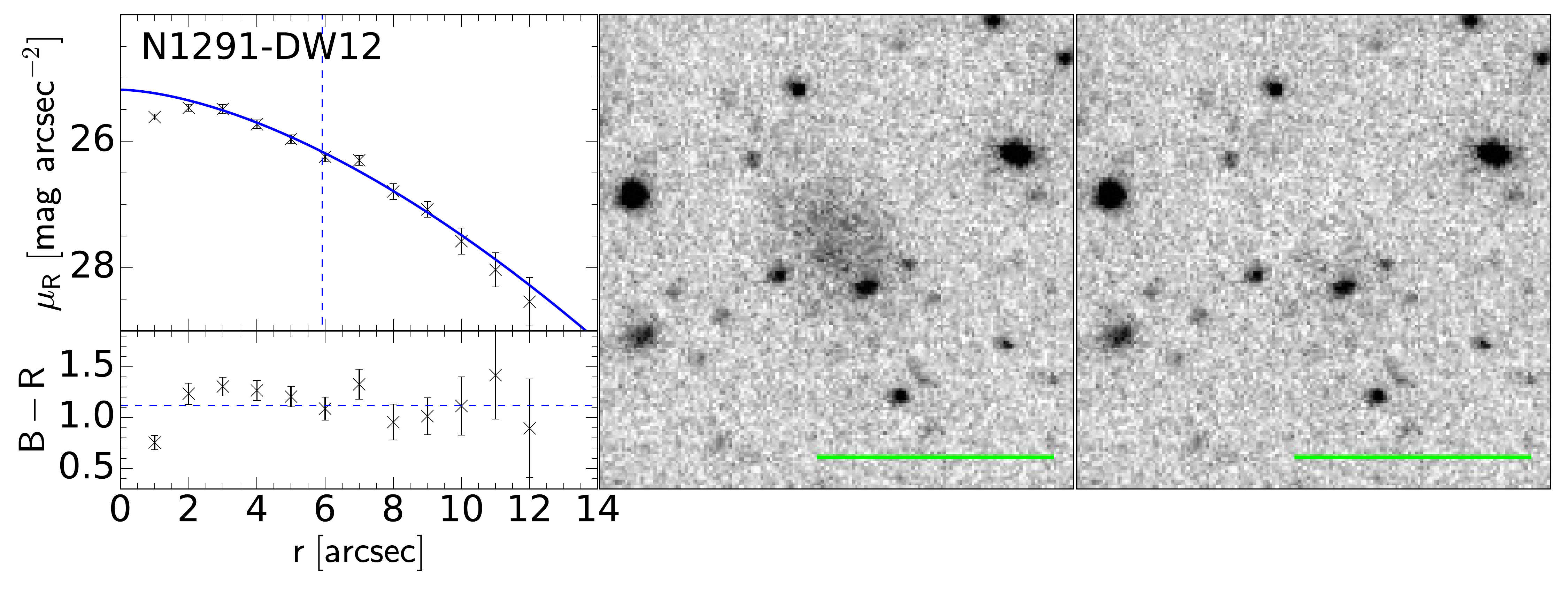}
\caption{(Continued)
} \label{fig:fig4}
\end{figure*}
\setcounter{figure}{3}

\begin{figure*}[]
\centering
\includegraphics[width=150mm]{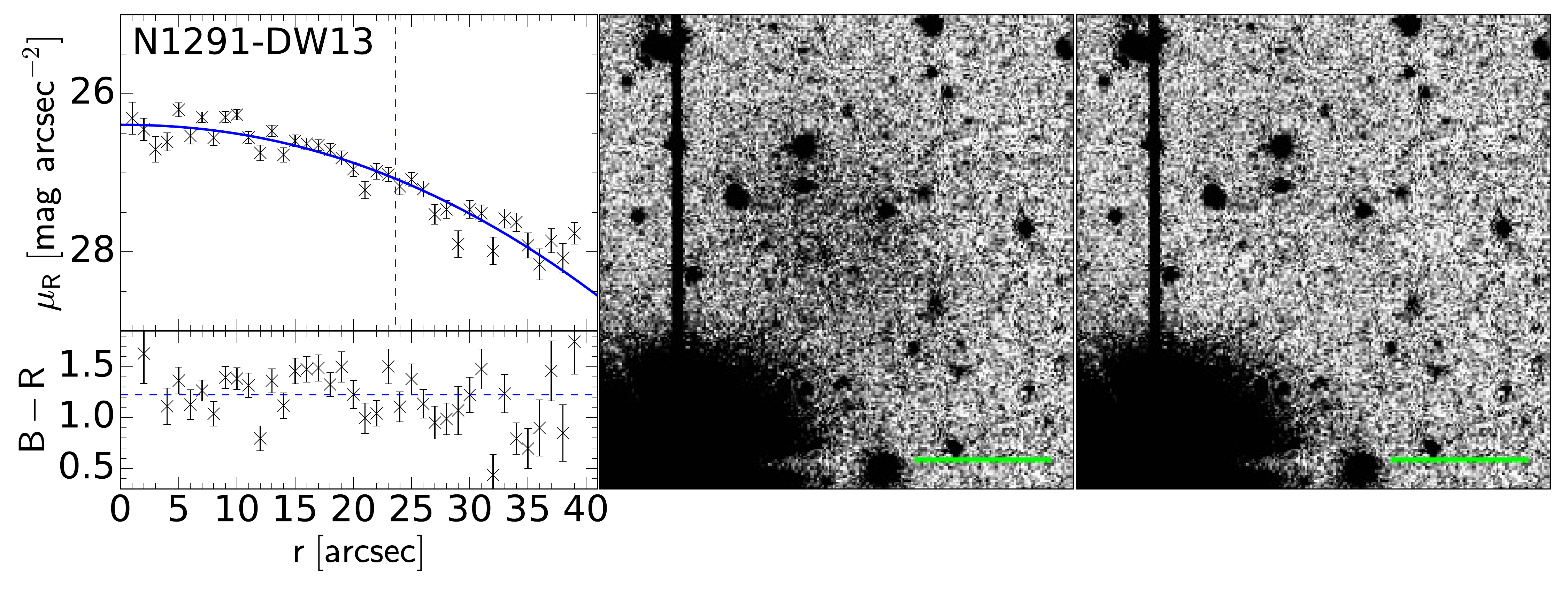}\vspace{-6mm}
\includegraphics[width=150mm]{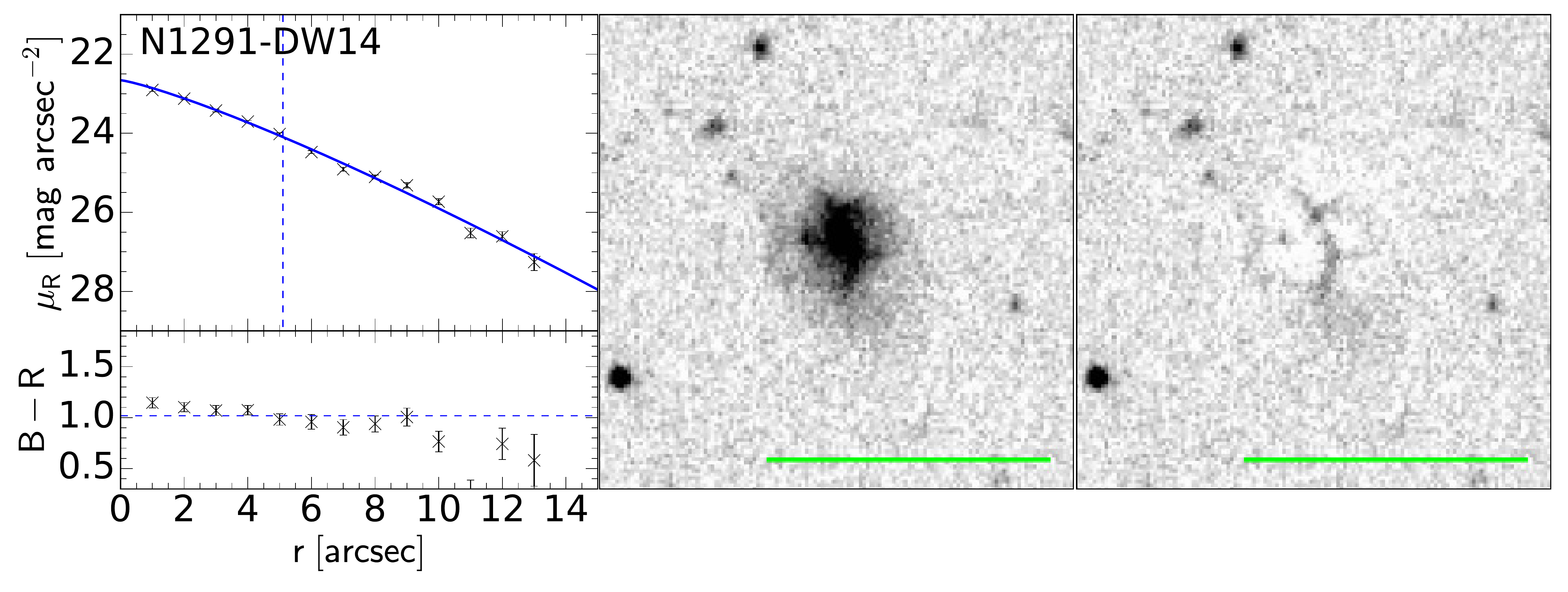}\vspace{-6mm}
\includegraphics[width=150mm]{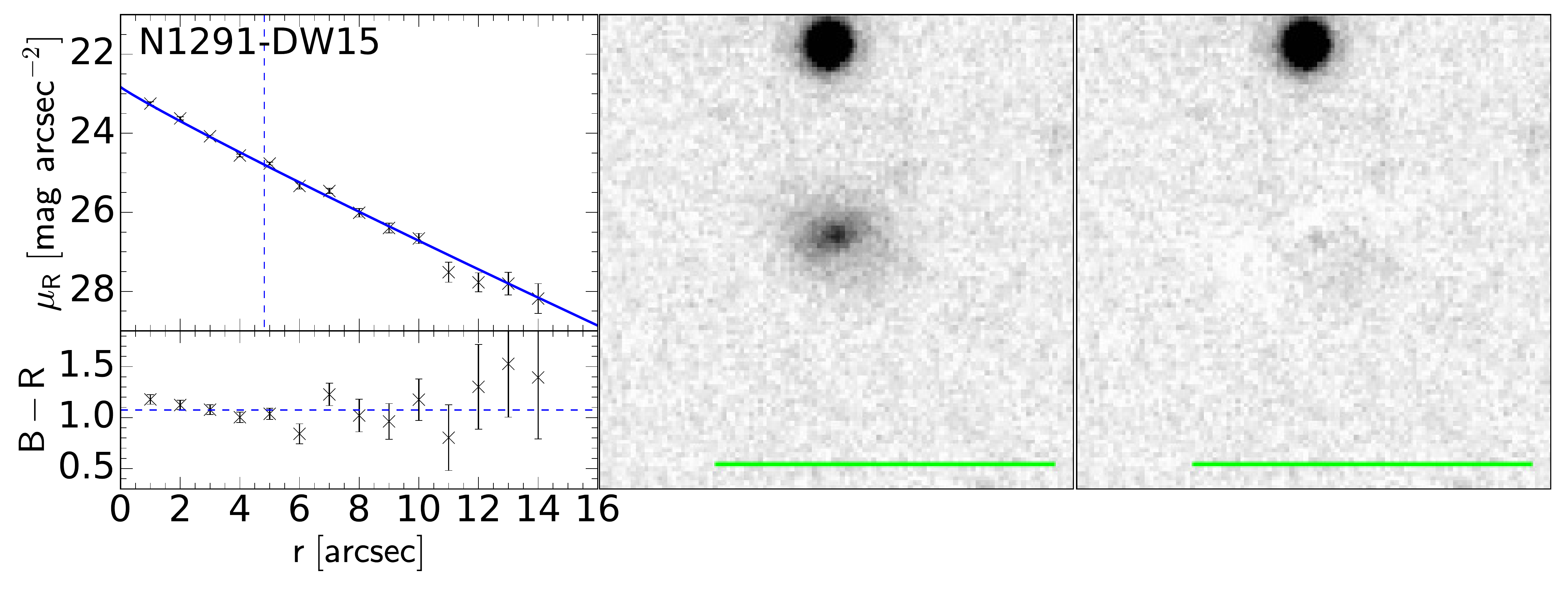}
\caption{(Continued)
} \label{fig:fig4}
\end{figure*}

\subsection{Structural and photometric properties}
This section presents the properties of the dwarf galaxy candidates mainly using 
$R$-band images (Table 2). 

The images were prepared by subtracting the sky background from each cutout image using a 
two-dimensional polynomial model. Note that the sky subtraction was performed before the mosaic images 
were generated, but there local sky fluctuations may have introduced residuals. The surface brightness profiles 
of the dwarf galaxy candidates were estimated in $B$- and $R$-band images 
using the {\tt ELLIPSE} task in IRAF. All parameters were first fit freely. Then the central position, ellipticity, 
and position angle values at the effective radius of each dwarf candidate were conservatively calculated 
based on the initial results. Finally, the isophotal fit was performed again using the three values above 
as fixed parameters over the whole radius. 

\begin{figure}[t]
\centering
\includegraphics[width=85mm]{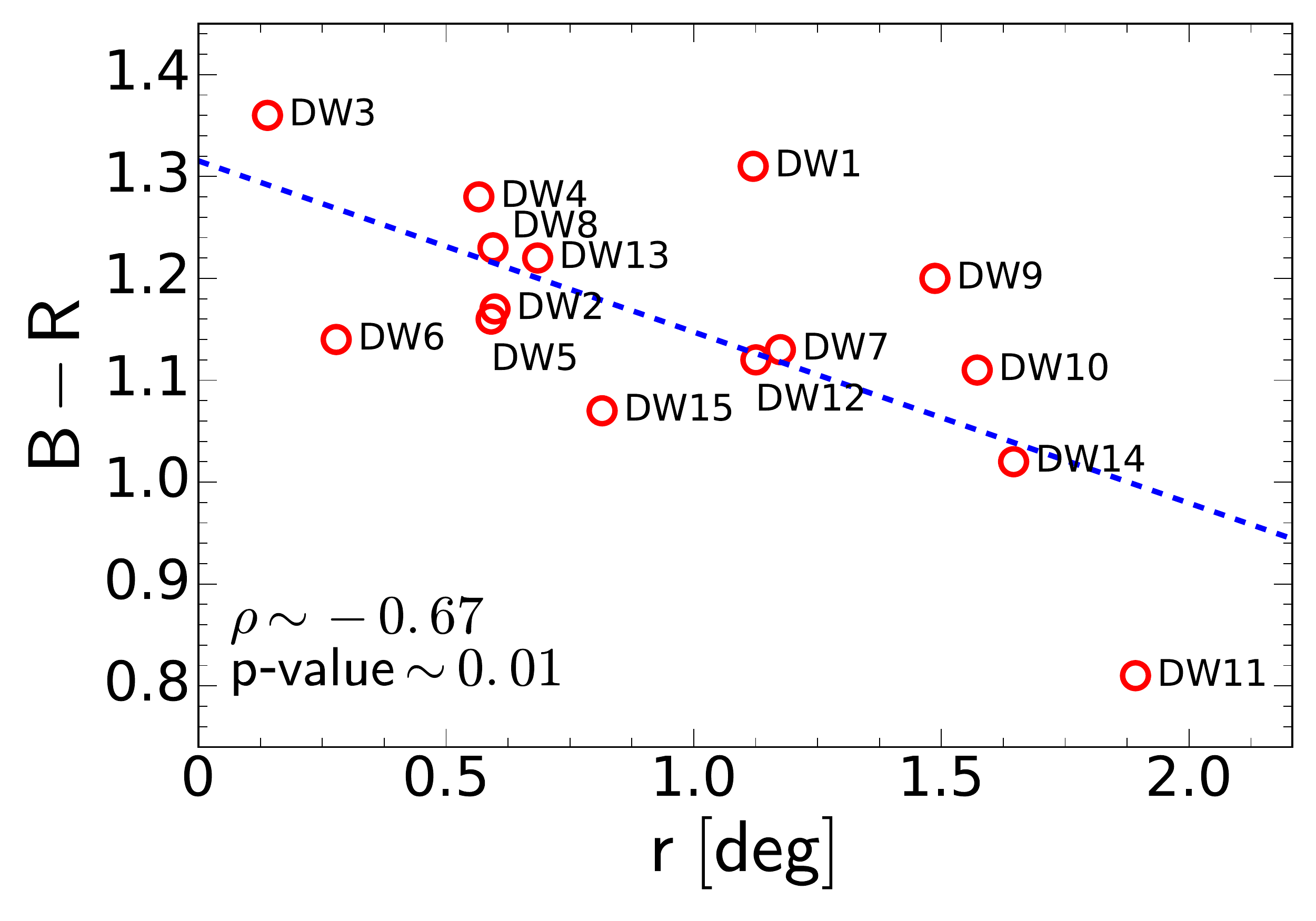}
\caption{$B-R$ color distribution of the dwarf candidates as a function of their projected distance from the center of 
NGC 1291. The dashed line shows a linear fit with 15 dwarf galaxy candidates. The correlation coefficient and 
p-value are noted at the bottom of the panel.
} \label{fig:fig5}
\end{figure}

Figure 4 shows both the original and model-subtracted cutout images. Most of the removed objects left no 
residuals, but some (N1291-DW7/DW14) left irregular features due to their clumpy structures. 
Note that they seem like background spiral galaxies, but their central surface brightnesses are somewhat 
fainter than that of the typical spiral galaxies ($\mu_{0,B} \sim 21.65$ \sbr; see \citealp{1993AJ....106.1394B} 
and references therein), implying that they are likely to be dwarf galaxies. Several dwarf spiral galaxies have 
been found in the nearby Universe although they are rare (cf. \citealt{1995AJ....110.2067S,2006AJ....132..497L}). However, without any 
confirmation of their distances, it is hard to determine the real nature of them. 

The azimuthally averaged surface brightness and color profiles are also presented in Figure 4. 
Note that the parameters presented in this section are not corrected by foreground Galactic extinction 
because the expected reddening effect towards NGC 1291 is negligible at $E(B-V)\sim0.01$. 

The central surface brightnesses of the candidates are distributed between $22.5 \lesssim \mu_{0,R}
\lesssim 26.5$ \sbr\ and their average central surface brightness is $\mu_{0,R}\sim$ 24.05 \sbr. 
The surface brightness profiles were fitted with a single S\'ersic function which is 
commonly expressed as 
\begin{equation}
I(r)=I_e exp\left\{-b_n\left[\left(\frac{r}{r_e}\right)^{1/n}-1\right]\right\},
\end{equation}
where $r_e$ is the effective radius, $I_e$ is the intensity at the effective radius, $n$ is the S\'ersic index, and 
$b_n = 1.9992n-0.3271$, a constant adopted from \citet{1989woga.conf..208C} 
(see also \citealp{1999A&A...352..447C}). 
Data points within the innermost 2 arcsec and which are fainter than 28 \sbr\ were not used for the fit. 
As a result, all dwarf galaxy candidates were well described with a single 
S\'ersic fit (blue solid lines in Figure 5). 
The effective radii were estimated to be $4.5 \lesssim r_e \lesssim 23.5$ 
arcsec (vertical lines in Figure 4). At a distance of 9.08 Mpc, it corresponds to $r_e\sim$ 200 pc to 1 kpc. 
S\'ersic indices were $0.5\lesssim n\lesssim1.0$ with an average 
value of 0.74. The total magnitudes of the dwarf galaxy candidates were estimated to be $15.5 
\lesssim m_R \lesssim 20.0$ mag using the following equation of
\begin{equation}
L_{tot}=2\pi I_0 {r_s}^2 n\Gamma (2n),
\end{equation}
where $I_0$ is the intensity at the center, $r_s$ is the scale length that satisfies the equation of 
$r_e\approx r_s(1.9992n-0.3271)^n$, and $\Gamma$ is the gamma function. 
Note that considering the image depth difference and large errors, there was no significant 
difference with the literature results of the four previously detected galaxies in terms of the central 
surface brightnesses and the apparent magnitudes. 

The weighted average colors are shown as horizontal dashed lines in Figure 4. 
The color range is 0.8 $\lesssim B-R \lesssim$ 1.5 mag and the average value is $\langle B-R\rangle\sim$ 
1.16 mag. Taking into account large errors in the outer regions, most of the candidates exhibited weak or 
no gradients in their color profiles. 

Furthermore, their stellar masses were calculated using the mass-to-light ratio of 
$\log(M_\star/L)=0.872\times(B-R)-0.866$ from exponential star formation history with Kroupa initial 
mass function developed by \citet{2013MNRAS.430.2715I}. The average stellar mass is 
$\sim$$1.35\times10^7$\solmass. The stellar masses were also estimated using other mass-to-light ratios 
found in \citet{2001ApJ...550..212B} to quantify the uncertainty of the stellar masses. 
The deviation for estimating stellar mass was revealed up to $\sim$20\%. 
Note that detailed stellar evolution uncertainty was not considered, so the true deviation may be much larger 
(cf. \citealt{2013ARA&A..51..393C}).

\begin{figure}[t]
\centering
\includegraphics[width=85mm]{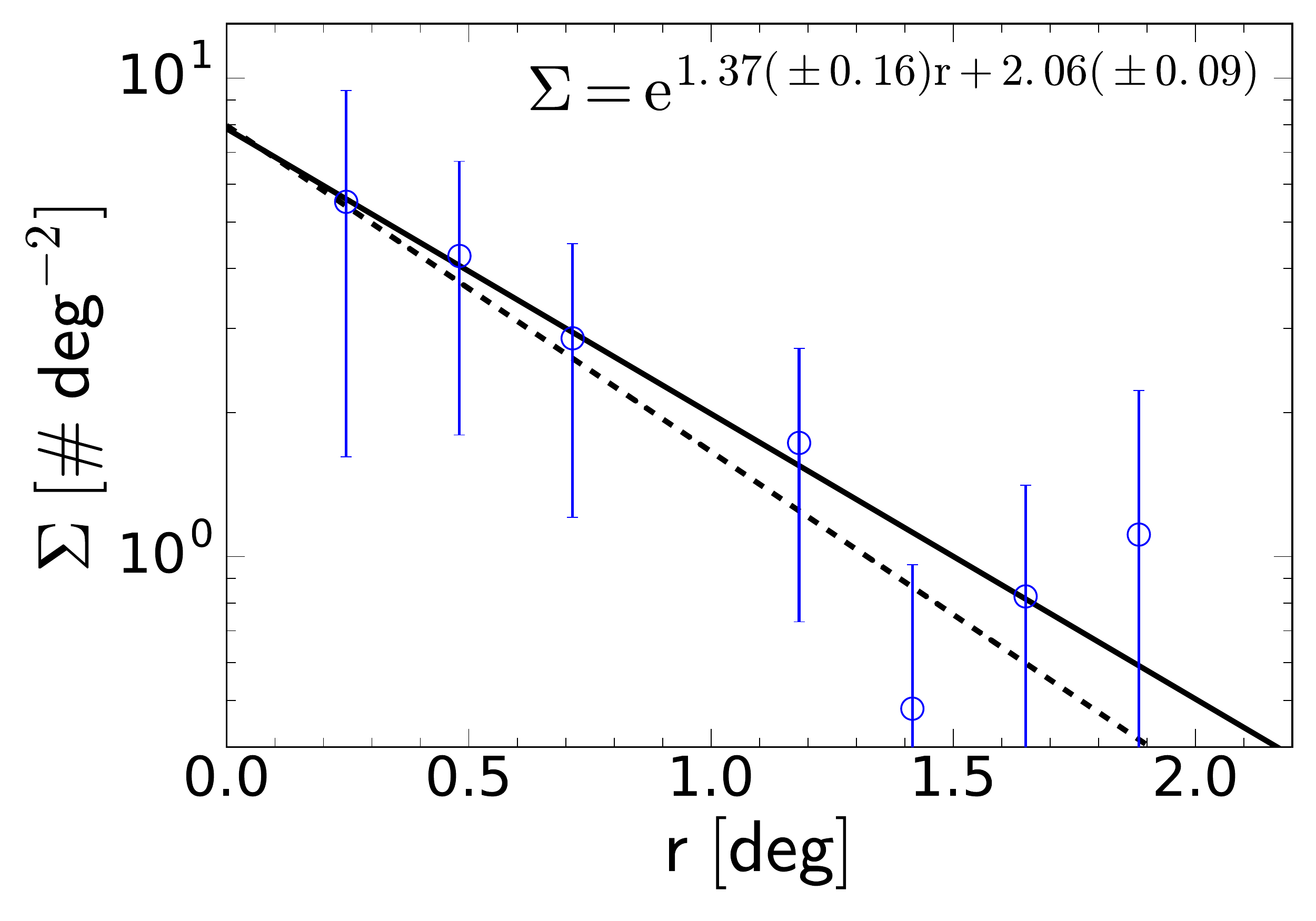}
\caption{The radial number density profiles of the dwarf galaxy candidates which show the results from 
two different binning methods: the equal linear distance binning (solid line) and the equal area binning (dashed line). 
The data points and the equation were derived from the former method.
} \label{fig:fig6}
\end{figure}

Figure 5 shows the $B-R$ color distribution of the dwarf galaxy candidates 
as a function of their projected distance from the center of NGC 1291. 
Though the candidates are significantly scattered, there seems to be a negative linear correlation, 
which has also been reported in much massive environments (e.g., \citealp{2019A&A...625A.143V}). 
Note that the color of dwarf galaxies can be affected by their masses and magnitudes 
(e.g., \citealp{1985AJ.....90.1681B,2002AJ....123.2246C,2006A&A...451.1159A,2012A&A...538A..69L,
2016ApJS..225...11Y}). We found that the color of the dwarf candidates are weakly correlated with the 
absolute magnitudes, but there is no statistically significant correlation 
between the total magnitudes and the projected distances. Indeed, \cite{2015MNRAS.447..698P} have 
reported that the quenching process of the satellite galaxies around isolated MW-like host galaxies can 
be effective at lower stellar masses ($<10^8$\solmass). Therefore, a negative color gradient for the 
projected distance can be interpreted as a result of the quenching process of the satellite galaxies in the 
halo of NGC 1291 which also implies that the dwarf candidates may be accompanying NGC 1291. 

Interestingly, N1291-DW3 and DW6 show weak excess of light in their innermost regions 
as nucleated cores. Given that they are the nearest candidates to NGC 1291, 
the nucleated core was likely a product of the tidal perturbation between the host and the satellite 
galaxies (cf. \citealp{1991A&A...252...27B,2000ApJ...543..620O}).
However, the exact mechanism behind nucleation in dwarf galaxies is still a subject of debate 
(cf. \citealt{2005MNRAS.363.1019G} and references therein). 
Moreover, \citet{2019ApJ...885...88P} mentioned that the environment in different dynamic states 
can affect the correlation between the nucleated fraction and other physical properties of dwarf galaxies. 
More surveys of dwarf galaxies must be conducted, especially of those in environments similar 
to the one found in the NGC 1291 system.

\subsection{Number density profile}
Figure 6 presents two different radial number density profiles for the dwarf galaxy candidates according to 
their projected distance from the center of NGC 1291. The small sample size made the fit sensitive to the 
binning size, so two binning sizes were used: (1) equal linear distance in each annulus and (2) equal 
area in each annulus. Each can be well described by an exponential function up to the virial radius of 
NGC 1291. The two methods yielded the number densities of $\Sigma = e^{1.37(\pm 0.16)r+2.06(\pm 0.09)}$ 
and $\Sigma=e^{1.57(\pm 0.56)r+2.07(\pm 0.38)}$, respectively. These results are relatively similar and 
within the error budget. It may be inappropriate to extract physically meaningful results from this fit 
because of the limited sample size on which it was based, but it is worth noting that 
the number density profiles showed centrally concentrated distributions, implying that the dwarf 
candidates are associated with NGC 1291.

\section{Discussion}

\begin{figure*}[t]
\centering
\includegraphics[width=120mm]{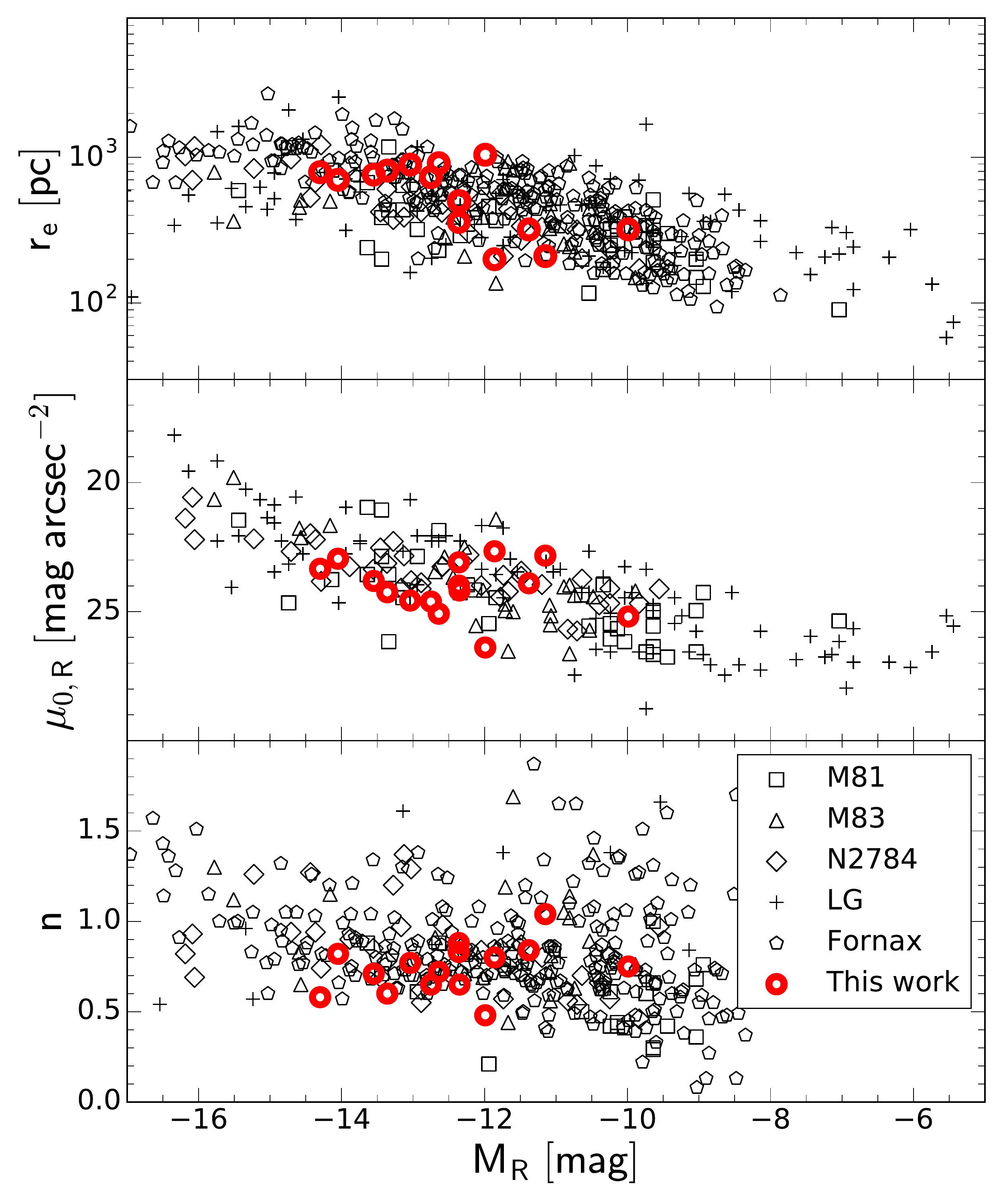}
\caption{The effective radii, central surface brightnesses, and S\'ersic 
indices of dwarf galaxies as a function of absolute magnitude. Red circles denote the NGC 1291 
candidates located at a distance of 9.08 Mpc. The squares, 
triangles, diamonds, crosses, and pentagons represent the dwarf galaxies in the M81 
group \citep{2009AJ....137.3009C}, the M83 group \citep{2015A&A...583A..79M}, 
the NGC 2784 group \citep{2017ApJ...848...19P}, the LG 
\citep{2000AJ....119..593J,2012AJ....144....4M}, and the Fornax cluster 
\citep{2015ApJ...813L..15M}, respectively. 
} \label{fig:fig7}
\vspace{5mm}
\end{figure*}

\subsection{Comparisons with dwarfs in other host systems}
NGC 1291 has been thought to reside in a relatively isolated environment. 
For instance, NGC 1291 accompanies a single bright galaxy ($M_R<-17$ mag) and 
15 dwarf galaxy candidates within the projected distance of $D_{proj}<300$ kpc. 
If the properties of dwarf galaxies are easily affected by the environment, the comparison 
between the dwarf candidates discovered in this study and those in other host systems with 
different populations can provide useful insight into the evolution of dwarf galaxies.
The following samples were chosen because they provided available data on 
structural and photometric properties of their satellites, regardless of their distance 
confirmations using resolved individual stars. 

Figure 7 presents the effective radii, central surface brightnesses,
and S\'ersic indices of the dwarf candidates as a function of absolute magnitude. 
For the sake of simplicity, we assumed that the dwarf candidates are located at the 
distance of NGC 1291 ($\sim$9.08 Mpc). The properties of known dwarfs in other 
host systems as well as the Fornax cluster are also presented. The structural and photometric 
properties of the dwarf candidates in the NGC 1291 system were revealed to be similar 
to those in other environments. Their significant distribution indicates that the 
shapes of dwarf galaxies are independent of their environments. This finding also supports the 
assumption that the dwarfs are located at approximately the same distances 
as NGC 1291. In fact, it cannot be a robust evidence as a shift in their distances would not entirely 
separate them from a distribution with the bulk of dwarf galaxies. However, most of the dwarf candidates 
are unlikely to be located in the foreground or background because no other host systems is found in the 
images that may be associated with dwarf candidates. 

\begin{figure}[t]
\centering
\includegraphics[width=85mm]{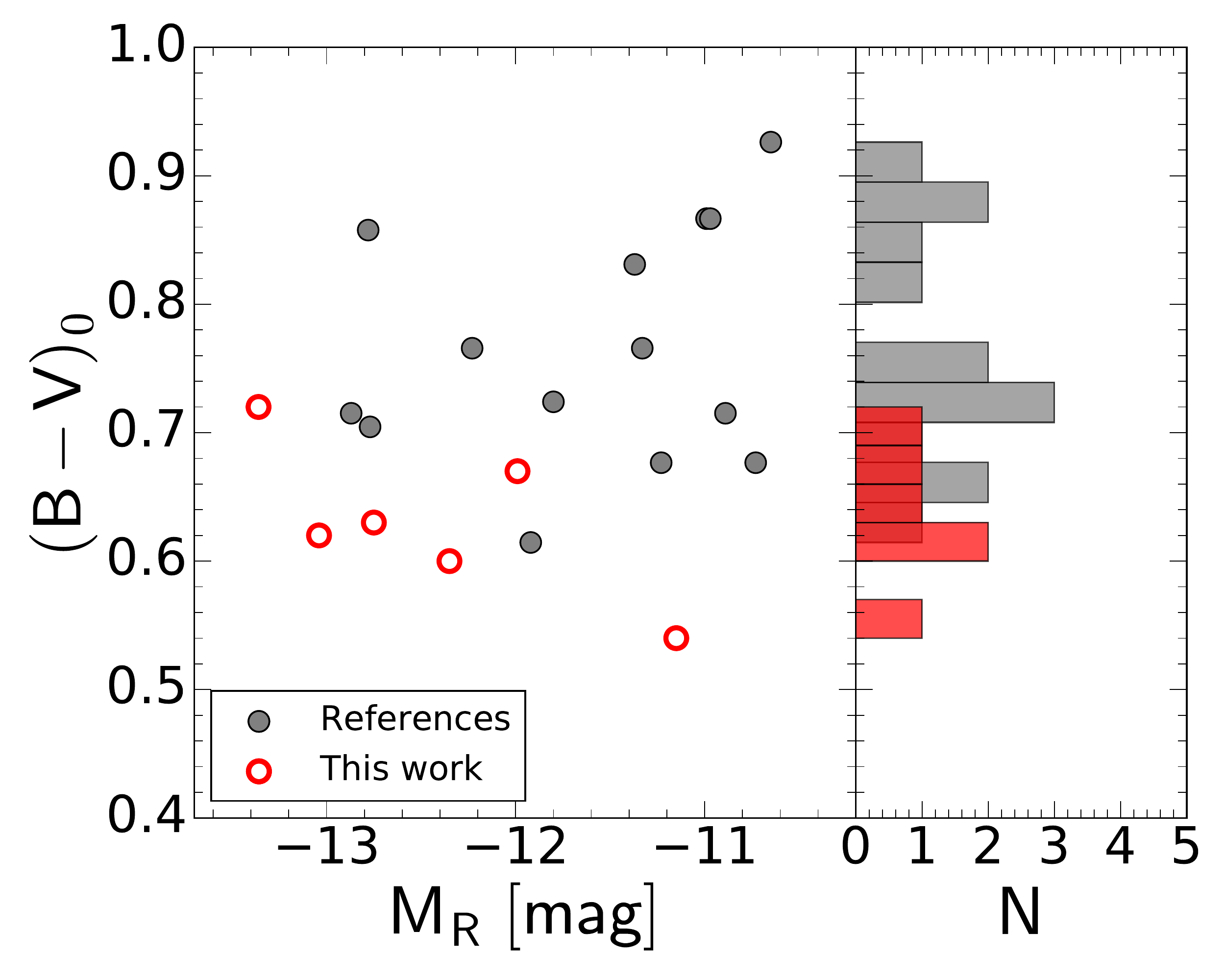}
\caption{The color distributions of dwarf candidates in the NGC 1291 system (red) and those of the 
M83, M101, M106, NGC 2784 groups (gray). The samples had to meet the criteria of  $D_{proj}\le 130$ 
kpc and $-13.5\le M_R\le -10.5$ mag.
} \label{fig:fig8}
\end{figure}

We also compared the colors of the dwarf candidates with those provided from the 
literatures. Interestingly, the average color of the dwarf galaxy candidates, 
($\langle B-V\rangle_0\sim0.63\pm0.10$)\footnote{We used the interpolated conversion 
factors of $V-R$, matching to the $B-R$ color of each candidate adopted from \citet{1995PASP..107..945F}}, 
is slightly smaller than that of the dwarfs in the M83 group ($0.82\pm 0.24$; \citealt{2015A&A...583A..79M}), 
the M101 group ($0.71\pm 0.04$; \citealt{2014ApJ...787L..37M}), the M106 group 
($0.70\pm0.19$; \citealt{2011MNRAS.412.1881K}), and the NGC 2784 group
($0.67\pm 0.17$; \citealt{2017ApJ...848...19P}). Note that the corrections for Galactic 
extinction were performed for all five samples and the $B-V$ colors in other studies were 
converted from $g$ and $r$-band using the equation of \citet{lupton05} if necessary. Indeed, this result may 
be inconclusive because of marginal uncertainties in color measurement, and systematic differences among 
the studies in the observational conditions. 
Nevertheless, we suggest that the comparison of color distributions under the strict conditions will be 
helpful in exploring the evolution of dwarf satellite galaxies in different host systems. 

The color of dwarf satellite galaxies can be affected by their own masses 
and their projected distances from their host galaxies, which as shown in Figure 5. 
Therefore, we utilized subsamples with the absolute magnitude range of $-13.5\le M_R\le -10.5$ 
mag to avoid the mass effect as well as the incomplete detection issue. Secondly, we used the projected 
distance cut with $D_{proj}\le 130$ kpc, adopted from the FoV limit of M106 survey 
\citep{2011MNRAS.412.1881K}. Note that the projected distance limit is broadly consistent with a half of the virial 
radius of each system, so dwarf galaxies within it are expected to have sufficiently experienced the environmental effect. 

\begin{figure*}[t]
\centering
\includegraphics[width=150mm]{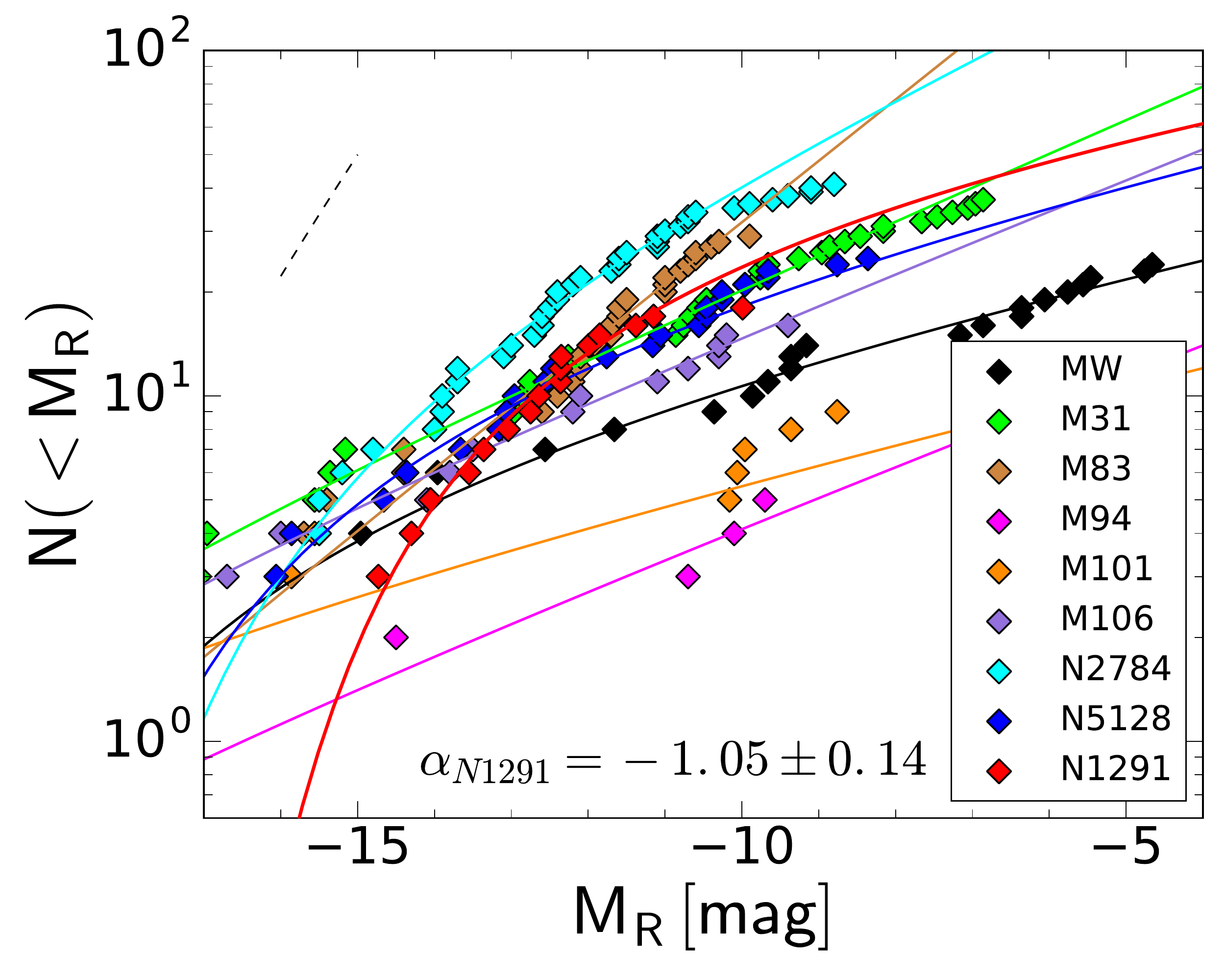}
\caption{The cumulative luminosity functions of satellite galaxies in the NGC 1291 system and 
several other host systems. Different colors denote different systems, 
as labeled in the plot. The solid lines show the fitted Schechter functions 
using galaxies in $M_R\gtrsim-17$ mag, which also have a completeness rate $>$90\%. 
The faint-end slope of the NGC 1291 system is noted at the bottom. The dashed line represents 
the faint-end slope predicted from the $\Lambda$CDM model.
} \label{fig:fig9}
\end{figure*}

Figure 8 shows the color distributions of dwarf candidates in the NGC 1291 system and other dwarf galaxies. 
Approximately 15 dwarfs in the M83, M101, M106, NGC 2784 groups were selected as counterparts 
and their average color is $\langle B-V\rangle_0\sim0.76\pm0.09$. Six of the dwarf candidates in the 
NGC 1291 system were selected with an average color of $\langle B-V\rangle_0\sim0.63\pm0.06$. 
The Kolmogorov-Smirnov test also indicated that the color distributions of the two samples are unlikely to 
have been drawn from the same population (p-value $\simeq0.01$). 
The dwarf candidates in the NGC 1291 system seem to have magnitude-dependent 
color distributions, while dwarf galaxies in other host systems tend to be redder regardless of their 
absolute magnitudes. This difference tends to be more obvious in less luminous dwarf galaxies which 
are thought to be susceptible to environmental effect due to their less massive dark matter halos. 
It shows that the dwarf satellite galaxies around NGC 1291 is somewhat less 
quenched than those in other system. It is noteworthy that although host galaxies resemble each 
other in terms of stellar mass, the evolution of dwarf satellite galaxies around them can vary. 

\subsection{Cumulative luminosity functions}
We examined the cumulative luminosity function (LF) of the NGC 1291 
system. The Schechter function \citep{1976ApJ...203..297S} was employed, which is: 
\begin{equation} 
N(<M)=\phi^*\Gamma[\alpha+1,10^{0.4(M^*-M)}],
\end{equation}
where $N(<M)$ is the number of galaxies brighter than $M$, $\phi^*$ is a scaling factor, 
$\alpha$ is the faint-end slope, $M^*$ is the characteristic magnitude where the break of 
LF occurs, and $\Gamma$ is the upper incomplete gamma function.
To improve the completeness of the LF, we included potential members of the NGC 1291 system 
using the radial velocity range of $839\pm 150$\kms\ (ESO 300-14; $M_R\simeq-17.79$, 
ESO 300-16; $M_R\simeq-14.73$). Their projected distances from the center of NGC 1291 
are $\simeq$230 kpc and $\simeq$280 kpc, respectively. Note that the velocity offset of 150\kms\ 
for NGC 1291 was defined based on the velocity distribution of nearby objects, but it did not 
significantly affect the slope of the faint-end.

Figure 9 compares the cumulative LF of the faint galaxies in the NGC 1291 system 
with those of the MW\&M31 \citep{2012AJ....144....4M,2013ApJ...772...15M,
2013ApJ...779L..10M}, M83 \citep{2015A&A...583A..79M}, M94 \citep{2013AJ....145..101K,
2018ApJ...863..152S}, M101 \citep{2019ApJ...885..153B}, 
M106 \citep{2011MNRAS.412.1881K}, NGC 2784 \citep{2017ApJ...848...19P}, 
NGC 5128 \citep{2019ApJ...872...80C}. Note that the incompleteness introduced from the 
different FoV covered by each study may affect the following results. 
Since the dwarf satellites in M83, M106, NGC 2784 and NGC 1291 have not yet been confirmed 
to be located at each host system, the resulting LFs can be regarded as upper limits at the moment.   
In order to directly compare the references, only galaxies with absolute magnitudes 
of $M_R\footnote{For the sake of simplicity, we used the conversion factor of $V-R=0.56$, 
$r-R=0.24$ from \citet{1995PASP..107..945F}} \gtrsim-17$ mag were used. 
The faint magnitude limits of the samples were adopted to have a completeness rate higher 
than 90\% in each study. We found that the faint-end slope in the NGC 1291 
system is $\alpha=-1.05\pm0.14$, which is much flatter than that predicted by the $\Lambda$CDM 
model ($\alpha=-1.8$; \citealp{2002MNRAS.335..712T}) shown as dashed line in Figure 9. 
It is similar to that of the MW ($-1.10\pm0.07$), M31 ($-1.24\pm0.13$), M83 ($-1.44\pm0.06$), 
M94 ($-1.22$), M101 ($-1.13$), M106 ($-1.22\pm0.17$), NGC 2784 ($-1.26\pm0.06$), NGC 5128 
($-1.11\pm0.18$). Note that we could not derive uncertainties for M94 and M101 because of their 
sparse satellite populations in the system. This result may indicate that the population of low-mass 
dwarf galaxies in the NGC 1291 system is indistinct from those in other host systems. 

\section{Summary}
In this study, 15 dwarf galaxy candidates were identified by visual inspection of deep and 
wide-field images of NGC 1291 obtained with KMTNet. Of those 15 candidates, 
11 are newly discovered in this study. We performed imaging simulations and estimated 
the completeness of the visual inspection that was above 70\% in $\mu_{0,R} 
\lesssim$ 26 \sbr\ and $M_R\lesssim-10$ mag. The photometric and structural properties 
of dwarf candidates were measured, the results of which are summarized as follows. 

\begin{enumerate}

\item{The surface brightness profiles of dwarf galaxy candidates were well described 
with a single S\'ersic function. The structural and photometric properties were similar to 
those of ordinary dwarf galaxies, with a central surface brightness of $\mu_{0,R}\sim$ 22.5--26.5 
\sbr\ and effective radii of 0.2--1 kpc. Assuming that they are located at 9.08 Mpc, the absolute 
magnitudes of the 15 dwarf galaxy candidates are $-14.5\lesssim M_R\lesssim -10$ mag.}

\item{Only the two nearest dwarf candidates from NGC 1291 exhibited nucleated cores. 
The colors of the dwarf candidates seem to be linearly correlated with their projected distances 
from the center of NGC 1291. This result may have been a product of the interaction between the 
host and satellite galaxies, such as tidal interactions and the quenching process.}

\item{In order to understand the evolution of dwarfs in different environments,
the properties of the dwarf candidates were compared to those in other host systems. 
The color distribution of dwarf candidates is the only property dissimilar from those of other dwarfs 
in other satellite population systems as the candidates are slightly bluer than other dwarfs. This result 
indicates that the dwarfs in the NGC 1291 system are likely less quenched than dwarfs in other host 
systems. It also reveals that environments play an important role in the quenching process but may 
not affect the morphology shaping of dwarf galaxies.}

\end{enumerate}

\acknowledgments
We are grateful to an anonymous referee for constructive comments and suggestions.
This research has made use of the KMTNet system operated by the Korea Astronomy and 
Space Science Institute (KASI) and the data were obtained at CTIO in Chile, one of three host sites. 
LCH was supported by the National Science Foundation of China (11721303, 11991052) and the 
National Key R\&D Program of China (2016YFA0400702). This research was supported by the 
National Research Foundation of Korea (NRF) grant funded by 
the Korea government (MSIP; NRF-2017R1C1B2002879). 
YKS acknowledges support from the National Research Foundation of Korea (NRF) grant 
funded by the Ministry of Science and ICT (NRF-2019R1C1C1010279). 
H.S.P. was supported in part by the National Research Foundation of Korea (NRF) grant funded by 
the Korea government (MSIT, Ministry of Science and ICT; NRF-2019R1F1A1058228).




\bibliography{ms}




\end{document}